\documentstyle[11pt]{report}
\begin{document}
\title{Hamilton's Formalism for Systems with Constraints}
\author{Andreas W. Wipf\\ \\
Institut f\"ur Theoretische Physik\\
Eidgen\"ossische Technische Hochschule\\
H\"onggerberg, CH-8093 Z\"urich, Switzerland}
\date{Lectures given at the Seminar "The Canonical Formalism in
Classical and Quantum General Relativity", Bad Honnef, September
1993\\
ETH-TH/93-48}
\maketitle
\begin{abstract}
The main goal of these lectures is to introduce and review
the Hamiltonian formalism for classical constrained systems
and in particular gauge theories. Emphasis is put on the relation
between local symmetries and constraints and on the relation
between Lagrangean and Hamiltonian symmetries.
\end{abstract}
\tableofcontents
\newcommand{\eqngr}[2]{\par\parbox{11cm}
{\begin{eqnarray*}#1\\#2\end{eqnarray*}}\hfill
\parbox{1cm}{\begin{eqnarray}\end{eqnarray}}}
\newcommand{\eqngrl}[3]{\par\parbox{11cm}
{\begin{eqnarray*}#1\\#2\end{eqnarray*}}\hfill
\parbox{1cm}{\begin{eqnarray}\end{eqnarray}}\label{#3}}
\newcommand{\eqngrrl}[4]{\par\parbox{11cm}
{\begin{eqnarray*}#1\\#2\\#3\end{eqnarray*}}\hfill
\parbox{1cm}{\begin{eqnarray}\end{eqnarray}}\label{#4}}
\def\kapha{{\kappa\over 2}}
\def\lap{\triangle}
\def\es{\!=\!}
\def\ms{\!-\!}
\def\vx{\vec{x}}
\def\qi{q^i}
\def\npi{p_i}
\def\dq{\dot{q}}
\def\ddq{\ddot{q}}
\def\dqi{\dot q^i}
\def\dpi{\dot p_i}
\def\ddqi{\ddot q^i}
\def\qti{q^{\tilde i}}
\def\dqti{\dot q^{\tilde i}}
\def\dpti{\dot p_{\tilde i}}
\def\ddqti{\ddot q^{\tilde i}}
\def\vy{\vec{y}}
\def\qj{q^j}
\def\pj{p_j}
\def\dqj{\dot q^j}
\def\dpj{\dot p_j}
\def\ddqj{\ddot q^j}
\def\qtj{q^{\tilde j}}
\def\ptj{p_{\tilde j}}
\def\dqtj{\dot q^{\tilde j}}
\def\dptj{\dot p_{\tilde j}}
\def\ddqtj{\ddot q^{\tilde j}}
\def\vph{\varphi}
\def\vpha{\vph^a}
\def\vphb{\vph^b}
\def\pa{\partial}
\def\vpham{\pa_\mu\vph^a}
\def\vphan{\pa_\nu\vph^a}
\def\phm{\phi_m}
\def\phn{\phi_n}
\def\phmpr{\phi_{m^\pr}}
\def\napprox{\approx\!\!\!\!\!/\,\,}
\def\Box{\Delta}
\def\al{\alpha}
\def\ka{\kappa}
\def\lam{\lambda}
\def\si{\sigma}
\def\eps{\epsilon}
\def\ov{\over}
\def\di{D\!\!\!\!\slash\,}
\def\fdi{\partial\!\!\!\slash\,}
\def\ddx{d^dx\,}
\def\ha{{1\over 2}}
\def\tr{\,{\rm tr}\,}
\def\pr{\prime}
\def\be{\bar\eta}
\def\ba{\bar\alpha}
\def\bpsi{\bar\psi}
\def\gam{\gamma}
\def\gan{\gamma_n}
\def\om{\omega}
\def\cd{{\cal D}}
\def\cl{{\cal L}}
\def\ch{{\cal H}}
\def\cs{{\cal S}}
\def\cm{{\cal M}}
\def\pclo#1{{\pa\cl\ov \pa {#1}}}
\def\plo#1{{\pa L\ov \pa {#1}}}
\def\pho#1{{\pa H\ov \pa {#1}}}
\def\pphmo#1{{\pa \phm\ov \pa {#1}}}
\def\>{\rangle}
\def\<{\langle}
\def\gf{\gamma_5}
\def\gff{\gamma_{d+1}}
\def\pan{\par\noindent}
\def\cov{\bigtriangledown}
\def\de{\delta}
\def\pd{\psi^{\dagger}}
\def\mtxt#1{\quad\hbox{{#1}}\quad}
\def\pamu{{\pa_\mu}}
\def\panu{{\pa_\nu}}
\def\al{\alpha}
\def\be{\beta}
\def\si{\sigma}
\def\gam{\gamma}
\def\pr{\prime}
\def\lam{\lambda}
\def\tr{\hbox{tr}\,}
\def\ta{{\tilde a}}
\def\tb{{\tilde b}}
\def\tc{{\tilde c}}
\def\tsi{{\tilde e}}
\def\tbe{{\tilde b}}
\def\tga{{\tilde c}}
\def\tde{{\tilde d}}
\def\ti{{\,\tilde i}}
\def\tj{{\,\tilde j}}
\def\hnn{\hat{0}}
\def\hr{\hat{R}}
\def\hi{\hat{I}}
\def\jm{\hat{N}}
\def\hio{\hat{I}_{\xi_1\lambda_1}}
\def\hit{\hat{I}_{\xi_2\lambda_2}}
\def\dxila{\delta_{\xi,\lambda}}
\def\dq{\dot{q}}
\def\cn{{\cal N}}
\def\tal{{\tilde a}}
\def\tbe{{\tilde b}}
\def\tga{{\tilde c}}
\def\tde{{\tilde d}}
\def\tsi{{\tilde e}}
\def\ctal{{C_\tal}}
\def\ctbe{{C_\tbe}}
\def\ctga{C_\tga}
\def\tttr{t^\tga_{\tal \tbe}}
\def\tttw{t^\tbe_\tal}
\def\str{f^c_{ab}}
\def\ha{{1\over 2}}
\def\una{{\vec A}}
\def\unb{{\vec B}}
\def\und{{\vec D}}
\def\unpi{{\vec \npi}}
\def\unpa{{\vec\pa}}
\def\qti{q^{\tilde i}}
\def\dqti{\dot q^{\tilde i}}
\def\dpti{\dot p_{\tilde i}}
\def\ddqti{\ddot q^{\tilde i}}
\def\qtj{q^{\tilde j}}
\def\ptj{p_{\tilde j}}
\def\dqtj{\dot q^{\tilde j}}
\def\dptj{\dot p_{\tilde j}}
\def\ddqtj{\ddot q^{\tilde j}}
\def\ti{{\,\tilde i}}
\def\tj{{\,\tilde j}}
\def\hr{\hat{R}}
\def\hi{\hat{I}}
\def\hio{\hat{I}_{\xi_1\lambda_1}}
\def\hit{\hat{I}_{\xi_2\lambda_2}}
\def\dxila{\delta_{\xi,\lambda}}
\def\tttr{t^\tga_{\tal \tbe}}
\def\tttw{t^\tbe_\tal}
\def\str{f^c_{ab}}
\newtheorem{satz}{Theorem}
\newtheorem{lemma}{Lemma}
\chapter{Introduction}
All fundamental field theories in physics
are invariant with respect to some group of
{\it local symmetry transformations}. For Yang-Mills theories
these are the gauge transformations, for string theory and
gravity space-time diffeomorphisms and
for supersymmetric theories local supersymmetry
transformations. In such theories which are called {\it gauge theories}
or more generally {\it singular systems,} the local symmetry
relates different solutions stemming from the same initial conditions
and the general solution of the
equations of motion contains arbitrary time-dependent functions.
Hence there is a continuous set of accelerations
which belong to the same initial position and velocity and
we expect that all accelerations correspond only to a subset
of initial conditions. This subset is defined by the Lagrangean
constraints\footnote{this statement is made more precise by the off-shell
Bianchi identities}. Hence all gauge theories are systems with
constraints. In the Hamiltonian formalism this means that
there are conditions on the allowed initial momenta and positions.
These conditions must then be conserved by the time evolution, and
this requirement may lead to further constraints.\par
One should distinguish between gauge theories with
internal symmetries and those which are {\it generally covariant}.\pan
For the former all constraints are linear in the momenta and
the Hamiltonian does not vanish. The local symmetry transformations
are generated by the first class constraints.\pan
For generally covariant theories at least one constraint
is quadratic in the momenta\footnote{topological field
theories are exceptional in this respect}
and there are canonical variables for which the Hamiltonian
$H$ itself is a constraint, usually called
{\it superhamiltonian}. This leads to the question whether
$H$ generates the dynamical time-evolution or
kinematical local symmetries as the other first class constraints.
We will hear more about this problem of interpreting time
in the lectures of R. Baig and P. Hajicek.\par
Attempts to handle constrained systems date back more than forty years.
In his classical works Dirac set up a formalism to treat such systems
self-consistently \cite{dir}. Later Bergman et.al. in a series of papers
investigated the connection between constraints and invariances
\cite{ab51,b56,b61}.
After the introduction of Grassmann variables to describe fermions
\cite{b66}, the formalism has been extended to include fields with half-integer
spins \cite{gs71,c76,bm77}. The development culminated with the advent
of the elegant and powerful BRST formalism \cite{brs76}.
These and other classical results have been
a prerequisite for the quantization of gauge theories
both in the path integral formalism \cite{f69,bf77}
and in the functional Schr\"odinger picture \cite{s83,j88,kw92}.
\par
Besides the classical lectures of Dirac \cite{d64} there are several
excellent reviews on the treatment of constrained systems.
Some focus more on systems with a finite number of degrees
of freedom \cite{sm74}, others on field theories \cite{hrt76} and some on
both \cite{s82,g91,ht92}. For generally covariant theories you may consult
\cite{gt90}.\par
In these hopefully selfcontained lectures I applied the developed
formalism at various stages to the abelian Chern-Simons model
with sources \cite{h84,p88,w89}. In this way the reader may become
acquainted with the constrained dynamics by way of example.
There are several aspects I could not cover in these lectures, especially
the inclusion of fermions which leads to graded Poisson-structures
\cite{gs71,c76} and the popular BRST formalism \cite{brs76}. Generally
covariant theories and in particular gravity are covered by the lectures of
R. Baig and N. Giulini on classical gravity and the other
contributions to the proceedings.\par
In the first part of these lectures I followed the conventional discussion
of constrained systems. The second chapter is devoted to
{\it singular Lagrangean systems}. I discuss the off-shell Bianchi
identities and show that all gauge theories are
constrained systems. In chapter three some important
facts about {\it constrained Hamiltonian systems} are reviewed and
discussed. In particular primary/secondary and first/second
class constraints, the generalized Legendre transformation
and the Dirac-Bergman algorithm are introduced.
The general formalism is applied to the {\it abelian Chern-Simons
theory} in chapter four. Then
we introduce the reduced phase space for first and second
class systems. Here the important Dirac brackets for second
class (SC) systems, the concept of observables and gauge
transformations for first class (FC) systems and the first order
formalism for mixed SC and FC systems are discussed.
Again I apply the general results to the
abelian Chern-Simons theory and show that
the only observables are the Wilson-loops.
In the last chapter I investigate the relation between
Lagrangean symmetries and the Hamiltonian gauge transformations
generated by the first class constraint \cite{mw92}. We shall see
that the latter must be supplemented by transformations which
vanish on-shell in order to recover the Lagrangean symmetries.
Also, I will discuss for which theories the equations of motion
follow from the local symmetries. Some of the results in
this last chapter are new and have not previously been published.
We feel that the results offered are somewhat novel.\par
I am indepted to V. Mukhanov for sharing with me many insights
and constributing to these notes.
\chapter{Singular Lagrangean systems}
\section{Singular Lagrangeans}
In these lectures I consider systems whose dynamics can be
derived from Hamilton's variational principle.
I assume that all Lagrangeans depend at most on first derivatives,
up to divergence terms. For higher derivative theories, and in particular
for higher derivative gravity, see \cite{gt90}. I use
local coordinates, unless I am forced to
address global questions, e.g. the Gribov problem or the role of
topologically nontrivial field configurations.
\par
With these assumptions the classical trajectories of a system with $N$
degrees of freedom make the action
\begin{equation}
S=\int\limits_{t_1}^{t_2} L(\qi ,\dqi)dt\mtxt{,}i=1,\dots,N
\end{equation}
stationary under variations $\delta q(t)$ which vanish at the endpoints.
The $q$ and $\dot q$ are local coordinates on the velocity phase
space $TQ$. The necessary conditions for $S$ to
be stationary are the {\it Euler-Lagrange equations}
\begin{equation}
L_i\equiv -{d\ov dt}\Big({\pa L\ov\pa\dqi}\Big) +{\pa L\ov \pa \qi}=0
\label{2.2}
\end{equation}
which can be rewritten as
\begin{equation}
L_i=-{\pa^2 L\ov \pa\dqi\pa\dqj}\ddqj-{\pa^2 L\ov\pa\dqi\pa\qj}\dqj
+{\pa L\ov \pa\qi}\equiv -W_{ij}(q,\dq )\ddqj+V_i=0.
\label{2.3}
\end{equation}
We see that the accelerations at a given time are
uniquely determined by $(q,\dq )$ at that
time only if the Hessian
$(W_{ij})$ can be inverted. Such systems are called {\it regular}.\par
If, on the other hand, $\det W\es 0$, the accelerations and
thus the time evolution will not
be uniquely fixed by the $(q,\dq )$. Such systems are called
{\it singular} \cite{ab51}. For singular systems different time evolutions
will stem from the same initial conditions.\par
The rank $R$ of $W$, which we assume for simplicity to be constant
on $TQ$, being $R\!<\!N$ implies the existence of $M\es N\ms R$
null-eigenvectors
\begin{equation}
Y^i_m (q,\dot q) W_{ij}(q,\dq)=0\mtxt{,}\quad m=1,\dots,M.
\end{equation}
Contracting the E-L equations \ref{2.3} with the $Y_m$ we get
\begin{equation}
\phm (q,\dq)\equiv Y^i_m V_i=0\mtxt{,}\quad m=1,\dots,M.
\end{equation}
These equations do not contain accelerations. Assume that
$M^\pr\!\leq \!M$ relations
\begin{equation}
\phmpr=0\mtxt{,} m^\pr =1,\dots,M^\pr,
\end{equation}
are functionally independent on the others, and the remaining ones
are either dependent or identically fulfilled. The independent
$\phmpr$ are the socalled {\it Lagrange constraints}.
\par
For {\it field theories} the dynamics is described by functions $\vpha(x)$
of spacetime with values in a certain target space.
The index $a$ may belong to an internal symmetry,
it may be a spacetime index or both internal
and spacetime index as in nonabelian gauge theories. When going
from point mechanics to field
theory one may think of replacing the disrete label $i$ by
a continuous one $(a,\vx )$:
\begin{equation}
q^i(t)=q(t,i)\longrightarrow q(t,a,\vx )=\vpha (t,\vx )=\vpha (x).
\end{equation}
Summations become integrals, e.g.
\begin{equation}
\sum\limits_i \dqi\dqi\longrightarrow
\sum\int d\vx\; \dot\vpha(\vx)\dot\vpha(\vx)
\end{equation}
and functions of $(q,\dq )$ become functionals of $\vph$ and $\dot\vph$.
Also, derivatives with respect to $\qi$ or $\dqi$ become functional
derivatives, e.g.
\begin{equation}
{\pa L\ov \pa\dqi}\longrightarrow
{\delta L\ov \delta \dot\vpha (\vx)}
\end{equation}
The suitable velocity phase space $TQ$ is chosen so that the
Lagrange-functional $L$ is continuous and sufficiently often differentiable.
If the target space is linear one may choose a Banach space
(typically a Sobolov space), otherwise one tries to model
the theory on a $C^k$-Banach manifold \cite{cm74,m74} since the implicit
function theorem still applies then.
Banach manifolds are modelled over Banach spaces and are
straightforward generalizations of finite-dimensional manifolds.\par
A functional on a Banach space $X$ is called continuous if
\begin{equation}
\lim_{n\to\infty }F[\vph_n]=F[\vph]\mtxt{for}X\ni\vph_n\to\vph.
\end{equation}
$F$ is called Frechet-differentiable at $\vph$ if there exists a linear
functional $F^\pr_\vph$ such that
\begin{equation}
\vert F[\vph+\delta\vph]-F[\vph]-F^\pr_\vph[\delta\vph]\vert
=o(\parallel \delta \vph\parallel )\mtxt{for all}
\parallel\delta\vph\parallel \to 0.
\end{equation}
For local theories the Lagrangean has the form
\begin{equation}
L[\vph,\dot\vph]=\int d\vx \,\cl(\vph,\pa_i\vph,\dot\vph)
\end{equation}
with a Lagrangean density $\cl$ depending only on the field and its
derivatives at the same point. For such theories the Euler-Lagrange
equations are
\begin{equation}
L_a\equiv -{\pa\ov\pa t}{\delta L\ov \delta\dot\vpha}+{\delta L\ov \delta\vpha}
=-\pa_\mu {\pa\cl \ov \pa(\pa_\mu\vpha)}+{\pa\cl \ov \pa\vpha}=0,\label{2.9}
\end{equation}
where I adopted the common notation
\begin{equation}
F^\pr_\vph[\delta\vph]\equiv \int{\delta F\ov\delta\vph (x)}\delta\vph (x).
\end{equation}
Rewriting the field equations as
\eqngrl{L_a&=&-{\pa^2\cl\ov \pa(\vpham)\pa(\pa_\nu\vphb)}\pa_\mu\pa_\nu\vphb
-{\pa^2\cl\ov \pa(\vpham)\pa\vphb}\pa_\mu\vphb+\pclo{\vpha}}
{&&\qquad\qquad\equiv -W_{ab}^{\mu\nu}\pa_\mu\pa_\nu\vphb+V_a=0}{2.10}
we can see that theories with $x^0$ taken as evolution parameter
are regular if $W_{ab}^{00}$ is invertible and singular if it is not.
For singular systems there exist (for each $\vx$) $M\es N\ms R$
null-vectors
\begin{equation}
Y^a_m(\vph,\pa\vph) W_{ab}^{00}(\vph,\pa\vph)=0\mtxt{,}
m=1,\dots,M
\end{equation}
which lead to nontrivial and independent relations
\begin{equation}
\phmpr (\vph,\pa\vph)\equiv Y^a_{m^\pr} V_a=0\mtxt{,} m^\pr=
1,\dots,M^\pr\leq M,
\end{equation}
involving only the fields and their first derivatives.
These are the Lagrangean constraints.\par
How one proceeds for singular systems is neatly explained
in \cite{s63,s82}. There are two points which have to be considered. Firstly
the rank of the Hessian may decrease if one takes the independent
constraints (2.6,12) into account.
This may lead to new independent constraints. Again the rank
may decrease leading to further constraints, etc. This process
terminates as soon as the rank does not change anymore.
\par
Secondly one needs to check whether the constraints one has found
after the above algebraic process has terminated are respected by the
{\it time evolution}. These may lead to new constraints. Again and
again differentiate newly emerging constraints until no new ones arise.
Add those relations involving accelerations to those already present.
Consistency of the old relations with the new ones may lead to further
constraints. After all that one needs again to check whether the rank of
the Hessian has changed. If this is the case  one needs to start from
the beginning etc.\pan
\section{Generalized Bianchi identities}
If a theory possesses a {\it local} gauge invariance we may map
solutions into solutions without affecting the initial conditions.
Thus we expect that {\it gauge theories are singular systems}.
Actually this follows from the generalized Bianchi identity
\cite{u59,t64} which we derive next.\par
The point transformations
\eqngr{x^\pr&=&x^\pr (x)\sim x+\delta x}
{\vph^\pr(x^\pr)&=&\vph^\pr (\vph(x),x)\sim\vph(x)+\delta\vph}
which leave the action invariant
\begin{equation}
\int d^dx^\pr\,\cl (\vph^\pr,\pa^\pr\vph^\pr,x^\pr)=
\int \ddx \cl (\vph,\pa\vph,x)
\label{2.14}
\end{equation}
form a group which we assume to be continuous.
For transformations close to the identity $d^dx^\pr\es d^dx
(1+\pa_\mu\delta x^\mu)$, and the invariance \ref{2.14} implies
\begin{equation}
\delta\cl+\cl\pa_\mu\delta x^\mu=\pa_\mu \lam^\mu
\end{equation}
with some $\lam$. Using
\begin{equation}
\delta\cl=\pclo{\vpha}\delta\vpha+\pclo{(\pa_\nu\vpha)}\delta(\pa_\nu\vpha)
+\pa_\mu\cl\delta x^\mu
\end{equation}
it follows at once that
\begin{equation}
\delta\cl+\cl\pa_\mu\delta x^\mu=
\pa_\mu(\cl\delta x^\mu)+\pa_\mu \big(\pclo{(\vpham)}\bar\delta\vpha\big)
+L_a\bar\delta\vpha,
\end{equation}
where the Euler derivatives $L_a$ have been defined in \ref{2.9} and
\begin{equation}
\bar\delta\vpha=\delta\vpha-\pa_\mu \vpha\delta x^\mu\sim
{\vpha}^\pr (x)-\vpha (x)
\end{equation}
is the infinitesimal difference of the old and the transformed
files at the same point. We used that $[\bar\delta,\pa_\mu]\es 0$.
Thus the gauge invariance implies
\begin{equation}
\fbox{$\displaystyle
\pa_\mu\big(\cl\delta x^\mu+\pclo{(\vpham)}\bar\delta\vpha-\lam^\mu\big)
+L_a\bar\delta\vpha=0$}\label{2.17}
\end{equation}
and these are the {\it generalized Bianchi identities}.
Nowhere did we use the equation of motion and thus \ref{2.17}
are {\it off-shell identities}. \par
First assume that $S$ is invariant under global transformations
forming a $n$-dimensional Lie-group. Then
\begin{equation}
\lam^\mu=\eps_\al\lam^{\al\mu},\quad\delta_\eps x^\mu
=\eps_\al A^{\al\mu},\quad
\delta_\eps \vpha=\eps_\al B^{\al a},\label{2.18}
\end{equation}
where the $\eps_\al,\al\es 1,\dots,n$ are the constant parameters of
the infinitesimal transformations. Inserting this into \ref{2.17} and
going on shell, $L_a\es 0$, we conclude
\begin{equation}
\pa j^{\al\mu}=0,\mtxt{where}
j^{\al\mu}=
\pclo{(\vpham)}\big(B^{\al a}-A^{\al\nu}\pa_\nu\vpha\big)
+\cl A^{\al\mu}-\lam^{\al\mu},\label{2.19}
\end{equation}
which {\it is Noether's first theorem}.
When deriving \ref{2.19} we imposed the
equations of motion so that the currents are {\it conserved only on-shell}.\par
Let us now assume that the symmetry transformations are local. In that case
the parameters become space-time dependent. Then \ref{2.18} generalizes to
\begin{equation}
\delta_\eps x^\mu=\eps_\al A^{\al\mu}\mtxt{,}
\delta_\eps \vpha=\eps_\al B^{\al a}+\pa_\mu \eps_{\al}C^{\al a\mu},
\end{equation}
where the $\eps_\al(x)$ parametrize the infinitesimal local gauge
transformations and $B$ and $C$ are the socalled {\it descriptors}
\cite{ab51}, which in general
depend on the fields and their derivatives. I assumed that no second
or higher derivatives of $\eps$ enter because this covers most interesting
examples. With
\begin{equation}
\bar\delta_\eps\vpha=\eps_\al\big(B^{\al a}-\vpham A^{\al\mu}\big)
+\pa_\mu\eps_\al C^{\al a\mu}
\end{equation}
the integrated form of \ref{2.17}, after a partial integration, reads
\begin{equation}
0=\int \eps_\al \big[L_a(B^{\al a}-\vpham A^{\al \mu})-\pa_\mu
(L_a C^{\al a\mu})\big].
\end{equation}
Since it must hold for arbitrary functions $\eps_\al$ this implies
that the expression between the square brackets must vanish.
Inserting $L_a$ from \ref{2.10} we end up with
\eqngrl{0&=L_a\big(B^{\al a}-\vpham A^{\al\mu}-\pa_\mu C^{\al a\mu}\big)
-C^{\al a\mu}\pa_\mu V_a}{&+C^{\al a\mu}\big(
\pa_\mu W^{\rho\sigma\rho}_{ab}\pa_\rho\pa_\sigma\vph^b
+W^{\rho\sigma}_{ab}\pa_\mu\pa_\rho\pa_\sigma\vph^b\big).}{2.22}
Since these are off-shell identities we conclude
\begin{equation}
C^{\al a(\mu}W^{\rho\sigma)}_{ab}=0,
\end{equation}
where the brackets around the indices mean symmetrization.
In particular, descriptors $C^{\al a 0}$ which are
not identically zero are null-eigenvectors of the Hessian,
\begin{equation}
C^{\al a 0}W^{00}_{ab}=0
\end{equation}
and render the system singular. If all $C^{\al a\mu}$ vanish,
then \ref{2.22} reduces to
\begin{equation}
0=\big(B^{\al a}-\pa_\rho\vpha A^{\al\rho}\big)L_a\Longrightarrow
\big(B^{\al a}-A^{\al\rho}\pa_\rho\vpha\big)W_{ab}^{(\mu\nu)}=0.
\end{equation}
Thus, if $C\equiv 0$ but the $B^{\al a}-A^{\al \rho}\pa_\rho\vpha$
are not identically zero, we conclude again that the system is singular.
So we have the important result that {\it gauge theories are necessarily
singular}. However, the converse is not true. Not all
conceivable singular systems are gauge theories.

\chapter{Hamilton's Formalism for Constraint Systems.}
\section{Primary constraints}
The departing point for the Hamiltonian formalism
is to define the canonical momenta (densities) by
\begin{equation}
p_i={\pa L\ov \pa\dqi }(q,\dq)\mtxt{resp.}
\pi_a(\vx )={\delta L\ov \delta\dot\varphi^a(\vx)},\label{3.1}
\end{equation}
where we assume that $L\in C^2(TQ)$. Only if
\begin{equation}
W_{ij}={\pa p_i\ov\pa\dq^j}\mtxt{resp.} W^{00}_{ab}(\vx,\vy)={\delta \pi_a(\vx)
\ov \delta \dot\varphi^b(\vy)}
\end{equation}
is invertible can this relation be solved for all velocities in terms
of the phase space variables, $\dq\es\dq(q,p)$ resp.
$\dot\vph=\dot\vph(\vph,\pi)$ \footnote{for field theories we assume
$TQ$ to be a Banach manifold so that the inverse function
theorem applies}. In the other case
not all momenta \ref{3.1} are independent, but there are some
relations
\begin{equation}
\phm (q,p)=0\mtxt{resp.}\phm (\vph,\pi)=0\mtxt{,}
m=1,\dots,M\label{3.2}
\end{equation}
that follow from the definition \ref{3.1} of the momenta.
In these lectures I assume that the constraints \ref{3.2} are independent.
\par
In the following we restrict ourselves to finite dimensional systems
and only comment on the related results for field theories.
The corresponding field theoretical formalae, if they apply,
are obtained if one uses deWitt's condensed notation \cite{dw65} in which $i$
becomes a composite index as explained above and in chapter 7.\par
The conditions \ref{3.2} are the $M\es N\!-\!R$ {\it primary constraints}.
They define the $2N\ms M$-dimensional {\it primary constraint surface},
denoted by $\Gamma_p$. The equations of motions have
not been used to derive them and they imply
no restriction on the $(q,\dq )$. \ref{3.1} maps the $2N$-dimensional
velocity phase space $TQ$ to the lower-dimensional submanifold
$\Gamma_p$ in the momentum phase space $\Gamma$. Hence the inverse
images of a given point in $\Gamma_p$ form a manifold of dimension
$M$.\par
To pass to the Hamiltonian formalism we impose some
{\it regularity conditions} on the primary constraints. They can
be alternatively formulated as:
\begin{enumerate}
\item the independent functions $\phm\, ,m=1,\dots,M$ can
be locally taken as the first $M$ coordinates of a new, regular,
coordinate system in the vicinity of $\Gamma_p$.
\item The gradients $d\phi_1,\dots,d\phi_M$ are locally
linearly independent on $\Gamma_p$; i.e., $d\phi_1\wedge\dots\wedge
d\phi_M\neq 0$ on $\Gamma_p$.
\end{enumerate}
For example, if $\phi$ is an admissible constraint, $\phi^2$ is not,
since $d(\phi^2)\es 2\phi d\phi\es 0$ on $\Gamma_p$. If the constraints
are regular the following properties hold.
\begin{satz}
If a smooth function $F(q,p)$ vanishes on $\Gamma_p$,
then $F=f^m\phi_m$ for some functions $f^m$.
\end{satz}
\begin{satz}If $\lam_i\delta q^i+\mu^i\delta p_i\es 0$ for arbitrary
variations $\delta q^i,\delta p_i$ tangent to the constraint surface,
then
\begin{equation}
\lam_i=u^m{\pa\phm\ov\pa q^i}\mtxt{and}
\mu^i=u^m{\pa\phm\ov\pa p_i}\quad\mtxt{on}\Gamma_p
\end{equation}
for some $u^m$.
\end{satz}
Before proving these two important theorems it is useful to distinguish
between weak and strong equations. A function $F(q,p)$ defined in
the neighbourhood of $\Gamma_p$ is called {\it weakly zero} if
\begin{equation}
F\vert_{\Gamma_p}=0\Longleftrightarrow F\approx 0
\end{equation}
and {\it strongly zero} if
\begin{equation}
F\vert_{\Gamma_p}=0\mtxt{and}
\big({\pa F\ov\pa \qi},{\pa F\ov \pa\npi}\big)\vert_{\Gamma_p=0}
\Longleftrightarrow F\simeq 0.
\end{equation}
These definitions are useful since the equations of motion
contain gradients of functions on $\Gamma_p$. The primary constraint
surface can itself be defined by weak equations. We have
\begin{equation}
\phm\approx 0\mtxt{but}\phm\simeq\!\!\!\!\!/\,\,0
\end{equation}
because of our regularity conditions on the constraints.\pan
Since $\nabla_x(f^m\phm)\approx f^m \nabla_x\phm$, where
$x\es (q,p)$ denotes the phase space coordinates, the first
theorem implies
\begin{lemma}
$F\approx 0\Longrightarrow F-f^m\phm\simeq 0$ for some
functions $f^m$.
\end{lemma}
To prove the first theorem we choose the independent constraints
$\phm$ as first coordinates of a regular coordinate system
$x\es(\phi ,\tilde x)$ in the neighbourhood of $\Gamma_p$. Since
$F(0,\tilde x)\es 0$ we have
\begin{equation}
F(\phi,\tilde x)=\int\limits_0^1{d\ov d\tau} F(\tau\phi,\tilde x)d\tau
=\phm\int\limits_0^1 F,_m(\tau\phi,\tilde x)d\tau
\end{equation}
and thus
\begin{equation}
F=f^m\phm\mtxt{with}f^m=\int\limits_0^1 F,_m(\tau\phi,\tilde x)d\tau.
\end{equation}
This proves theorem 1 in the neighbourhood $U$ of any
point on $\Gamma_p$. We cover the neighbourhood
of $\Gamma_p$ by open sets $U_i$, on each of which theorem 1 applies.
Together with the open sets $V_k$ on which $\phi_k\neq 0$ the
$U_i$ cover the whole phase space. On $V_k$ we can set $F=(F/\phi_k)\phi_k$
and theorem 1 holds there. Finally, to guarantee that the $f^m$
are the same on the overlap of $U_i$ and $U_{i^\pr}$ one
uses a finite partition of unity.\par
Theorem 2 follows immediately from the regularity condition
which implies that at a given point $x$ on $\Gamma_p$
a basis of $T_x\Gamma_p$ (the vectors tangent to $\Gamma_p$
at $x$), together with the gradients
$\nabla_x\phm$ form a basis of $T_x\Gamma$.
The assumption in theorem 2 means that
$(\lam,\mu)$ are orthogonal to $T_x\Gamma_p$. Thus it must
be a linear combination of the gradients $\nabla_x\phm$.\par
For {\it field theories} one finds
\begin{equation}
F[\phi,\tilde x]\approx 0\Rightarrow F=\int \,f^m\phi_m
\mtxt{,}f^m(\vx)=\int d\tau{\delta F\ov \delta\phi_m(\vx)}
[\tau\phi,\tilde x]
\end{equation}
and a weakly vanishing functional is a linear combination
of smeared constraints. The testfunctions should lie in the
space dual to the space of the constraints \cite{b77}.

\section{Legendre transformation}
The canonical Hamiltonian
\begin{equation}
H=\dqi\npi-L\mtxt{resp.} H=\int d\vx\;\pi_a(\vx)\dot\vpha(\vx)-L=
\int d\vx\; \ch
\end{equation}
has the remarkable property that $\dq$ enters $H$
only through the combination $p(q,\dq )$. This follows from
\eqngrl{\delta H&=&\dqi\delta\npi+\delta\dqi\npi-\delta\dqi\plo{\dqi}
-\delta\qi\plo{\qi}}{&=&\dqi\delta\npi-\delta\qi\plo{\qi}}{3.5}
which shows that $H$ is a function of $p$ and $q$ only.
Here $\delta p$ is to be regarded as linear combination of
$\delta q$ and $\delta\dq$ so that $\delta q,\delta p$
are tangent to $\Gamma_p$. $H$ is only defined on $\Gamma_p$ since
we used the constraints. We would like to extend
the formalism to the whole phase space $\Gamma$.\par
The equation \ref{3.5} can be rewritten as
\begin{equation}
\big(\pho{\qi}+\plo{\qi}\big)\delta\qi+\big(\pho{\npi}-\dqi\big)
\delta\npi=0\label{3.6}
\end{equation}
with variations tangent to $\Gamma_p$. $H$ may be the restriction
to the hypersurface $\Gamma_p$ of a function $\tilde H$ defined
all over phase space. Then \ref{3.6} holds with $H$ replaced by
$\tilde H$. Applying theorem 2 we conclude that
\eqngrl{\dqi&\approx& {\pa\tilde H\ov \pa\npi}+u^m \pphmo{\npi}}{
-\plo{\qi}&\approx&{\pa\tilde H\ov \pa\qi}+u^m\pphmo{\qi}.}{3.7}
The first set of relations enables us to recover the velocities from
the $(q,p)\in\Gamma_p$ and the parameters $u^m$. Because
of the regularity conditions on the constraints two
different $u$ yield different $\dq$ and
the first relation permits us to express $u$ as function
of $q$ and $\dq$. This way one obtains an {\it invertible
Legendre transformation} from the $2N$-dimensional velocity phase space to
the $2N$ dimensional space $\Gamma_p\times \{u^m\}$:
\begin{equation}
\npi=\plo{\dqi}(q,\dq )\mtxt{and} u^m=u^m(q,\dq )
\end{equation}
with inverse transformation
\begin{equation}
\dqi={\pa\tilde H\ov \pa p_i}+u^m\pphmo{\npi}\mtxt{and}\phm (q,p)=0.
\end{equation}
We had to extend the
Hamiltonian, which was originally defined only on $\Gamma_p$,
to a neighbourhood of $\Gamma_p$. According to theorem 1
two possible extensions differ by
a term $c^m\phm$. Thus the formalism should be unchanged by the
replacement
\begin{equation}
\tilde H\longrightarrow \tilde H+c^m(q,p)\phm.
\end{equation}
Indeed, making this transformation in \ref{3.7} just shifts the
$u$ to $u+c$.\par
Finally, the relations \ref{3.7} allow us to rewrite the equation of motion
\ref{2.2} in the equivalent Hamiltonian form
\begin{equation}
\dqi\approx\pho{\npi}+u^m\pphmo{\npi}\mtxt{and}
\dpi\approx-\pho{\qi}-u^m\pphmo{\qi},\label{3.9}
\end{equation}
where we dropped the tilde atop $H$.
The Lagrangean equations of motion \ref{2.2} are equivalent to \ref{3.9}.
The phase space function
\begin{equation}
H_p\equiv H+u^m\phm
\end{equation}
os the {\it primary Hamiltonian}. \par
Introducing the Poisson bracket of two phase space functions
\eqngr{
\{F,G\}&\equiv& {\pa F\ov \pa \qi}{\pa G\ov \pa\npi}-{\pa F\ov\pa\npi}
{\pa G\ov \pa\qi}\quad \mtxt{resp.}}{
\{F,G\}&\equiv& \int d\vx\Big({\delta F\ov \delta \vpha(\vx)}{\delta G\ov\delta
\pi_a(\vx)}-{\delta F\ov\delta\pi_a(\vx)}{\delta G\ov \delta\vpha(\vx)}\Big)}
and using $u^m\nabla_x\phm\approx \nabla_x(u^m\phm)$, the
Hamiltonian equations of motion can be rewritten as
\eqngrl{
\dqi&\approx \{\qi,H_p\}\approx\{\qi,H\}+\{\qi,\phm\}u^m}{
\dpi&\approx \{\npi,H_p\}\approx\{\npi,H\}+\{\npi,\phm\}u^m.}{3.12a}
Besides there are still the equations defining $\Gamma_p$:
\begin{equation}
\phm (q,p)=0.\label{3.12b}
\end{equation}
For an any phase-space function $F(q,p)$ the {\it time evolution}
follows then from
\begin{equation}
\fbox{$\displaystyle
\dot F\approx\{F,H_p\}\approx \{F,H\}+u^m\{F,\phm\}. $}
\label{3.13}\end{equation}
\section{Dirac-Bergman algorithm}
As in the Lagrangean formalism the constraints must be consistent
with the time evolution. If initially $(q,p)$
is on $\Gamma_p$ it should remain there at later
times. This means that the equations of motion should
preserve the constraints and
this gives rise to the {\it consistency conditions} \cite{dir,ab51}
\begin{equation}
\dot\phm\approx\{\phm,H\}+\{\phm,\phn\}u^n\equiv
h_m+C_{mn}u^n\approx 0.\label{3.14}
\end{equation}
For non-admissable Lagrangeans these relations will be inconsistent.
As an example take $L\es \dq-q$ which leads to $H\es q$ and $\phi\es p\ms 1$
so that \ref{3.14} reads $1\approx 0$. For such inconsistent models
the action has no stationary points and we shall exclude them.\par
To discuss the consistency relations \ref{3.14} we distinguish the two
following cases:\vskip .2truecm\pan
i) $\det C\napprox 0$:\pan
In this case $u$ is uniquely fixed by \ref{3.14} to be
$u^n\approx C^{nm}h_m$, where $C^{nm}$ is the inverse of $C_{nm}$.
The time evolution \ref{3.13} of a phase space function becomes
\begin{equation}
\dot F\approx \{F,H\}-\{F,\phm\}C^{mn}\{\phn,H\}.\label{3.15}
\end{equation}
No additional conditions appear. For any initial data $(q,p)$
on $\Gamma_p$ the time evolution stemming from \ref{3.15} is unambiguous and
stays on $\Gamma_p$.\vskip.2truecm\pan
ii) $\det C\approx 0$:\pan
In this case $u$ is not fixed and \ref{3.14} is only solvable if
$h_m w^m_a\approx 0$ for all left null-eigenvectors $w_a$
of $C$. Either these equations are
fulfilled or they lead to a certain number $K_1$ of new constraints
\begin{equation}
\phi_k\approx 0\mtxt{,}k=M+1,\dots,M+K_1\equiv J_1,
\end{equation}
called {\it secondary constraints}. The primary and
secondary constraints $\phi_j\approx 0,\,j=1,\dots,J_1$
define a hypersurface $\Gamma_1\subseteq \Gamma_p$.\pan
Now one has to check consistency for the primary and newly generated
secondary constraints on $\Gamma_1$,
\begin{equation}
\dot\phi_j=\{\phi_j,H\}+\{\phi_j,\phn\}u^n\equiv
h_j+C_{jn}u^n=0\mtxt{on}\Gamma_
1
\end{equation}
with the rectangular $J_1\times M$ matrix $C$.
The left null-eigenvectors $w^j_a$ of $C_{jn}$ imply further
conditions $w^j_a h_j\es 0$ on $\Gamma_1$ and
may lead to further, socalled {\it tertiary
constraints} which, together with the primary and secondary constraints,
define a hypersurface $\Gamma_2\subseteq \Gamma_1$, etc.\par
This procedure terminates after a finite number of iterations
and the following situation is reached:
There is a hypersurface $\Gamma_c\subset \Gamma$ defined by
\begin{equation}
\phi_j\approx 0\mtxt{,}j=1,\dots,M+K\equiv J.
\end{equation}
For every left null-eigenvector $w^j_a$ of the rectangular matrix
$C_{jm}=\{\phi_j,\phm\}$ the conditions $w^j_a\{\phi_j,H\}\approx 0$
are fulfilled. For the multiplier fields there are the equations
\begin{equation}
\{\phi_j,H\}+\{\phi_j,\phm\}u^m\approx 0,\label{3.18}
\end{equation}
where $\approx$ now means equality on $\Gamma_c$.
We note that the primary constraints are merely consequences
of the definition of the momenta, whereas we used the equations
of motion to arrive at the secondary constraints\footnote{in the
sequel I call all non-primary constraints secondary}.
\par
We make the same regularity assumptions on the full set of constraints
$\phi_j$ defining $\Gamma_c$ as we made on the primary constraints $\phm$
defining $\Gamma_p$. Also, we assume that the rank of $C$ is constant on
$\Gamma_c$.\par
\section{First and second class constraints}
The distinction between primary and secondary constraints will be
of minor importance in the final form of the Hamiltonian theory.
A different classification of contraints, namely into first and
second class \cite{d64}, will play a central part.
Let $v_a$ be a basis of the kernel of $C$,
\begin{equation}
\{\phi_j,\phm\}v^m_a\approx 0\mtxt{,}a=1,\dots,
\hbox{dim\ Ker}\,C=M-\hbox{rank}\,C.
\end{equation}
The general solution for the multipliers $u$ in \ref{3.18} has then
the form
\begin{equation}
u=\tilde u +\mu^a v_a,\label{3.20}
\end{equation}
where $\tilde u$ is a particular solution. We have separated
the part of $u$ that remains undetermined by the consistency
conditions. This part contains $M\ms\hbox{rank}\,C$ free functions
$\mu^a$.\pan
The combinations of primary contraints
\begin{equation}
\phi_a=v^m_a\phm
\end{equation}
weakly commute with all other constraints,
\begin{equation}
\{\phi_a,\phi_j\}\approx 0\mtxt{,}j=1,\dots,J.
\end{equation}
Moreover, since the $v^a$ form a base of Ker$\,C$, the $\phi_a$
are a complete set of primary contraints with this
property. This leads to the concept of {\it first class functions}
and in particular {\it first class constraints} (FCC). A function
$F(q,p)$ is said to be first class if its Poisson brackets with
{\it all constraints} vanish weakly (on $\Gamma_c$),
\begin{equation}
\{F,\phi_j\}\approx 0\mtxt{,}j=1,\dots,J.
\end{equation}
The set of first class functions is closed under Poisson brackets \cite{d64}.
This is proved as follows: if $F,G$ are first class, then according
to theorem 1
\begin{equation}
\{F,\phi\}=\phi^\pr\mtxt{,}
\{G,\phi\}=\phi^{\pr\pr}
\end{equation}
for any constraint $\phi$, where $\phi^\pr,\phi^{\pr\pr}$ are
some linear combinations of the constraints.
Using the Jacobi identity we have
\begin{equation}
\{\{F,G\},\phi\}=\{F,\{G,\phi\}\}-\{G,\{F,\phi\}\}=\{F,\phi^{\pr\pr}\}-
\{G,\phi^\pr\}\approx 0.
\end{equation}
In particular the constraints $\phi_a$ are a complete set of
first class primary constraints (modulo squares of second class
constraints). Also, as a result of the Dirac-Bergman
algorithm $H_p$ is first class.\par
A function that is not first class is called {\it second class}.
I use a notation adapted to this new classification.
All primary and secondary FCC are denoted by $\gam_a$.
The remaining constraints are called second class constraints (SCC)
and I denote them by $\chi_\al$.\par
The first property we need is that the matrix of SCC
\begin{equation}
\Delta_{\al\beta}=\{\chi_\al,\chi_\beta\}
\end{equation}
is {\it non-singular}. Indeed, if it was singular, then there would
exist a null vector $r^\al\Delta_{\al\beta}\approx\{r^\al\chi_\al,
\chi_\beta\}\approx 0$. Since $r^\al\chi_\al$ also commutes weakly with
the FCC (by their first class property)
it would weakly commute with all constraints and would be
first class which contradicts our assumption. For counting degrees of
freedom it is important to note that the number of SCC must be even.
Otherwise the antisymmetric $\Delta$ would be singular.
\par
Now consider the consistency conditions \ref{3.18}.
They are identically fulfilled for the $\gam_a$.
For the SCC we have
\begin{equation}
\{\chi_\al,H\}+\Delta_{\al\beta}u^\beta\approx 0,
\end{equation}
where $u^\beta\es 0$ if $\chi^\beta$ is a secondary SCC.
Solving for the multipliers we obtain
\begin{equation}
\Delta^{\beta\al}\{\chi_\al,H\}=\Big\{{u^\beta,\quad\ \chi_\beta
\mtxt{primary}\atop \quad 0,\qquad\chi_\beta\mtxt{secondary},}
\end{equation}
where $\Delta^{\al\beta}\Delta_{\beta\gamma}\es \delta^\al_{\;\gamma}$.
Thus all multipliers belonging to the primary SCC
are determined by the consistency conditions
and we remain with the undermined multipliers $\mu^a$ in \ref{3.20}.
We have the important result that the {\it number of undetermined
multipliers is equal to the number of independent primary FCC}.\par
Inserting that into the equations of motion \ref{3.13} we end up with
\begin{equation}
\dot F\approx \{F,H\}+\{F,\phi_a\}\mu^a-\{F,\chi_\al\}\Delta^{\al\beta}
\{\chi_\beta,H\},\label{3.25}
\end{equation}
where the $\phi_a$ are the primary FCC.
One can easily check that all constraints are preserved in time.

\chapter{Abelian Chern-Simons Theory with Sources}
To see how the general formalism works in an explicit
example I consider the abelian Chern-Simons model \cite{h84,p90,il92,fs93}.
This is a field theory for a gauge potential $A_a$
in $3$ space-time dimensions with coordinates $x\es (x^0,x^1,x^2)
\equiv (t,\vx )$ with first order Lagrangean density
\eqngrl{
\cl&={\kappa\ov 4} A^a\eps_{abc}F^{cb}+A^a J_a,\mtxt{where}}{
F_{ab}&=\pa_aA_b-\pa_bA_a\;\;\mtxt{,}\;\;\pa^a J_a=0.}{4.1}
Indices are lowered with the metric $\eta_{ab}=\hbox{diag}(1,-1,-1)$
and $\eps_{abc}$ is the Levi-Civita symbol, $\eps_{012}\es 1$.
We enclose the system in a finite box $[0,L]\times [0,L]$.
The quantum theory is sensitive to the value of the coupling constant
$\kappa$. For rational $2\pi\kappa$ and vanishing external current
$J$ the Hilberspace becomes finite-dimensional \cite{p90}.\par
For arbitrary periodic currents the action is invariant under
$U(1)$-gauge transformations
\begin{equation}
A^a\to A^a+\pa^a\lam\mtxt{,}
S\to S+\oint n^a\big({\kappa\ov 4}\lam\eps_{abc}F^{cb}+\lam J_a\big)
\label{4.2}
\end{equation}
provided $\lambda$ vanishes at the initial and final times and
$\lam,F_{01},F_{02}$ are periodic in $x^1,x^2$ with period $L$.
So we shall {\it assume} that $\lam$ and $F_{ab}$ are both periodic.\par
Since $\cl$ is linear in the first derivatives the Hessian vanishes
identically and the model is singular. Thus we expect $3$ independent
primary constraints (per space point). More explicitly, the canonical
momentum densities are
\begin{equation}
\pi_a(\vx)={\delta L\ov \delta\dot A^a (\vx)}=\kapha A^b(\vx)\eps_{ba0}
\end{equation}
and immediately lead to the {\it primary constraints}
\begin{equation}
\{\phm\}=\{\pi_0,\pi_1+\kapha A^2,\pi_2-\kapha A^1\}
\mtxt{,}m=1,2,3.
\end{equation}
The canonical Hamiltonian becomes
\eqngr{H&=&\int d\vx \Big(\pi_a(\vx)\dot A^a(\vx)-\cl \Big)=\int d\vx\,\ch}{
&=&-\int d\vx \Big(\kapha A^a\eps_{abi}\pa^i A^b+A^a J_a\Big)\mtxt{,}i=1,2.}
and the time evolution is determined by \ref{3.13} with primary Hamiltonian
\begin{equation}
H_p=\int d\vx\,\ch_p\mtxt{,}\ch_p=\ch+u^m\phm,
\end{equation}
and fundamental Poisson bracket
\begin{equation}
\{A^a(\vx),\pi_b(\vy)\}=\delta^a_{\;b}\delta(\vx-\vy).
\end{equation}
Let us now see whether secondary constraints arise from the
consistency conditions $\dot\phm\approx 0$. One computes
\begin{eqnarray}
\dot\phi_1(\vx)&=&\int d\vy\;\{\pi_0(\vx),\ch_p (\vy)\}\nonumber\\
&=&-\int d\vy\big\{\pi_0(\vx ),\kappa A^0(\vy)\eps_{0ji}\pa^i A^j(\vy)
+J_0(\vy)A^0(\vy)\big\}\nonumber\\
&=&\int\Big(\kappa\eps_{0ji}\pa^i A^j(\vy)+J_0(\vy)\Big)\delta(\vx-\vy)
\nonumber\\&=&\eps_{0ji}\kappa \pa^i A^j(\vx)+J_0(\vx)\nonumber
\end{eqnarray}
leading to the {\it secondary constraint}
\begin{equation}
\phi_4(\vx)=\kappa F_{12}-J_0(\vx).
\end{equation}
Of course there is a quicker way to arrive at this conclusion,
since $\dot\phi_1=-\pa\ch_p /\pa A^0$.\par
The time derivative of the other two primary constraints are
\eqngrl{\dot\phi_2(\vx)&=&\kappa\big(u^3-\pa^2 A^0(\vx)\big)+J_1(\vx)}{
\dot\phi_3(\vx)&=&\kappa\big(-u^2+\pa^1 A^0(\vx)\big)+J_2(\vx)}{4.8}
and putting them weakly to zero fixes the multipliers $u^2,u^3$.
Finally, we must have
\begin{equation}
\dot\phi_4(\vx)=\kappa\pa^2u^2(\vx)-\kappa\pa^1u^3(\vx)+\pa_0 J_0(\vx)
\approx 0.
\end{equation}
Inserting $u^2,u^3$ from \ref{4.8} this becomes $\pa_aj^a\es 0$ and
yields no further condition. Thus the Dirac-Bergman algorithm
leads to $3$ primary and $1$ secondary constraint.\par
Obviously $\phi_1\equiv\gam_1$ is first class. Also the combination
\begin{equation}
\gamma_2=\pa_1 \phi_2+\pa_2 \phi_3+\phi_4=\pa_i\pi_i
+\kapha F_{12}-J_0
\end{equation}
is first class and is the analog of the Gauss constraint in
electrodynamics. As SCC we may take
\begin{equation}
\chi_1=\phi_2\mtxt{and}\chi_2=\phi_3\Longrightarrow
\Delta_{\al\beta}=\kappa\pmatrix{0&1\cr-1&0\cr}\delta(\vx-\vy).
\end{equation}
$\Delta$ has inverse $\Delta^{-1}\es-\Delta/\kappa$ and \ref{3.25} reads
\begin{equation}
\dot F\approx \{F,H\}+\{F,\gam_1\} u^1+
{1\ov\kappa}\int d\vy\{F,\chi_i(\vy)\}\eps_{ij}\{\chi_j(\vy),H\}.
\end{equation}
Since the FCC commute with all constraints
they generate transformations on $\Gamma_c$, i.e.
if $(A,\pi)$ is on $\Gamma_c$ then
\eqngr{\delta_\cn \,A(\vx)&=\int d\vy\; \{A (\vx),\gam_a(\vy)\}\;\cn^a(\vy)}{
\delta_\cn \,\pi(\vx)&=\int d\vy\; \{\pi (\vx),\gam_a(\vy)\}\;\cn^a(\vy)}
are variations tangent to it. This follows from
$\delta_{\cn}\phi_j=\int\{\phi_j,\gam_a\}\cn^a\approx 0$. Also, since
these transformations commute with $H_p$, one
expects that they are related to infinitesimal gauge transformations.
Indeed, defining
\begin{equation}
G=\int d\vy \Big(\pa_0\lam(\vy)\gam_1(\vy)+\lam(\vy)\gam_2(\vy)\Big)
\end{equation}
one finds
\begin{equation}
\delta_\lam A^a(\vx)=\pa^a\lam(\vx).
\end{equation}
Only a particular combination of the $2$ FCC generate
the gauge transformations \ref{4.2}. \par
Contrary to the gauge generator
$G$ the primary Hamiltonian contains only the primary FCC.
A symmetric treatment of all FCC
is achieved in the extended formalism \cite{d64} discussed below.
We will come back to the important connection between FCC
and gauge transformations in chap.7.

\chapter{The reduced phase space.}
First we shall consider SC systems for which
no multipliers remain in the time evolution \ref{3.25}.
There is no ambiguity in the dynamics. The
term in \ref{3.25} containing the inverse of $\Delta$ forces
the system to stay on $\Gamma_c$. This
surface will be the {\it reduced phase space} for
SC systems.
\par
In the second subsection we consider FC systems. These are the
most important systems since all gauge theories are of this
type. Gauge related
point should be identified and this leads us to the
problem of gauge invariant functions and/or the gauge
fixing problem. The FCC together with a complete set of
gauge fixing conditions form then a SC system.
Thus for FC systems the gauge fixing define a subset
$\Gamma_r\in\Gamma_c$ and this set is the reduced phase space.
\section{Second class constraints and Dirac bracket}
Motivated by \ref{3.25} one introduces the {\it Dirac bracket}
\cite{dir} for two phase space function as
\begin{equation}
\fbox{$\displaystyle
\{F,G\}^*\equiv \{F,G\}-\{F,\chi_\al\}\Delta^{\al\beta}\{\chi_\beta,G\},
$}\label{5.1}
\end{equation}
in terms of which
\begin{equation}
\dot F\approx \{F,H\}^*
\end{equation}
for SC systems.
This bracket possess the same properties as the Poisson bracket,
i.e. they are antisymmetric, bilinear and obey the Jacobi identity
and product rule. In addition we have
\begin{equation}
\{F,\chi_\al\}^*=0\mtxt{,}
\{F,G\}^*\approx\{F,G\}\mtxt{,}
\{F,\{G,K\}^*\}^*\approx\{F,\{G,K\}\}
\end{equation}
for arbitrary $F$ and first class $G,K$. These properties follow easily
from the definition \ref{5.1} and the property that first class functions
have vanishing Poisson bracket with all constraints, e.g.
\begin{equation}
\{F,\chi_\al\}^*=\{F,\chi_\al\}-\{F,\chi_\beta\}\Delta^{\beta\gamma}
\{\chi_\gam,\chi_\al\}=0.\label{5.4}
\end{equation}
Let us draw an immediate consequence of \ref{5.4}. According to theorem 1
any function can be replaced by its restriction to $\Gamma_c$, up to
a linear combination of the constraints. Thus when calculating the
Dirac bracket \ref{5.1} between two functions we may replace them
by their restriction to $\Gamma_c$ since the other brackets vanish
on account of \ref{5.4}. It follows that the SCC
can be set equal to zero either before or after evaluating the
Dirac bracket. \par
To understand the {\it geometric meaning} of SCC we recall some
facts from symplectic geometry \cite{a78}:\par
In most cases the phase space $\Gamma$ is the cotangental bundle
$T^*Q$ over the configuration space $Q$ and hence is equipped with
a natural symplectic structure (a non-degenerate closed two-form)
\begin{equation}
\om=\om_{\mu\nu}{dx^\mu\wedge dx^\nu}
\end{equation}
which, according to Darboux, can locally be written as
\begin{equation}
\om=dq^i\wedge dp_i.
\end{equation}
Given $\om$ we can assign to a functions its corresponding Hamiltonian
vector field as
\begin{equation}
F\Longrightarrow X_F\mtxt{by} i_{X_F}\om=dF,
\end{equation}
where $i_X$ and $d$ are the interior and exterior derivatives,
respectively. In local coordinates we find
\begin{equation}
\om_{\mu\nu}X_F^\mu Y^\nu=\om(X_F,Y)=i_{X_F}\om(Y)=dF(Y)=\pa_\nu F\,Y^\nu
\end{equation}
for any vector field $Y$, or
\begin{equation}
X_F^\mu=-\om^{\mu\nu}\pa_\nu F, \mtxt{where}\om^{\mu\nu}\om_{\nu\rho}=
\delta^\mu_{\;\rho}.
\end{equation}
The Poisson bracket of two functions is
\begin{equation}
\{F,G\}=-\pa_\rho F\om^{\rho\sigma}\pa_\sigma G=\om_{\mu\nu}
\om^{\mu\rho}\pa_\rho F\om^{\nu\sigma}\pa_\sigma G=
\om(X_F,X_G).\label{5.8}
\end{equation}
In particular, the change of $F$ under the Hamiltonian flow generated
by $G$ can be written as
\begin{equation}
F^\pr\equiv \{F,G\}=\om (X_F,X_G)=i_{X_F}\om(X_G)=dF(X_G)=
X^\mu_G\pa_\mu F.\label{5.9}
\end{equation}
In other words, the flows generated by $G$
are just the motions along the Hamiltonian vector field $X_G$.
For $G\es H$ \ref{5.9} are the Hamiltonian equations of motion.\par
Finally there is an important relation between Poisson
and Lie bracket,
\begin{equation}
[X_F,X_G]=-X_{\{F,G\}},\mtxt{where} [X,Y]^\mu=X^\al\pa_\al Y^\mu-
Y^\al\pa_\al Y^\mu\label{5.10}
\end{equation}
are the Lie bracket. \ref{5.10} is a consequence of the
Jacobi identity.\par
Let us now return to the SC systems. The inclusion map
$j:\Gamma_c\longrightarrow \Gamma$ induces a two-form on $\Gamma_c$,
namely the pull back of the symplectic form $\om$ on $\Gamma$,
$\om_c=j^*\om $. $\om_c$ is closed since $\om$ has
this property, but it may be degenerate. In this case
it is called {\it pre-symplectic}. However, for SCC it is indeed symplectic,
as follows from
\begin{satz}
The $\chi_\al$ are second class if and only if
$\om_c=j^*\om$ is non-degenerate.
\end{satz}
Actually, we shall see that the Dirac bracket belongs to $\om_c$.
Most properties of the Dirac bracket, and in particular the
Jacobi identity follow then at once from the corresponding
properties of $\om$.\par
To prove this theorem we must show that $\om$ is non-degenerate
on the vectors tangent to $\Gamma_c$. A vector field $Y$ is tangent
to $\Gamma_c$ if $Y^\mu\pa_\mu\chi_\al$ vanishes for all constraints
$\chi_\al$. With \ref{5.9} this is equivalent to
\begin{equation}
\om(X_\al,Y)\approx 0\mtxt{for all} X_\al\equiv X_{\chi_\al}.\label{5.11}
\end{equation}
On the other hand, from \ref{5.8} follows that
\begin{equation}
\om (X_\al,X_\beta)=\{\chi_\al,\chi_\beta\}\equiv \Delta_{\al\beta}
\napprox 0
\end{equation}
so that the Hamiltonian vector fields generated by the constraints
are not tangent. Let us now determine the vectors $X$ which obey
\begin{equation}
\om(X,Y)\approx 0\mtxt{for all tangent }Y.\label{5.13}
\end{equation}
Since $\om$ is non-degenerate \ref{5.13} can have dim$\Gamma$-dim$\Gamma_c$
independent solutions $X$. But because of our regularity conditions
on the constraints the dim$\Gamma$-dim$\Gamma_c$ Hamiltonian vector fields
$X_\al$, which are not tangent, are independent solutions. Thus any $X$
which obeys \ref{5.13} is a combination of the $X_\al$. Hence
there cannot be a tangent $X$ obeying \ref{5.13} which proves that $j^*\om$
is non-degenerate.\par
Note that we used the SC nature of the constraints and in particular
that the {\it flows generated by the SCC
lead off the constraint surface}.\par
Now it is easy to prove that the Dirac bracket furnishes an
explicit representation for the induced Poisson bracket. For that
consider
\begin{equation}
\{F,G\}^*=\om(X_F,X_G)-\om(X_F,X_\al)\Delta^{\al\beta}\om(X_\beta,X_G)
\equiv\om^*(X_F,X_G).\label{5.14}
\end{equation}
It is easy to see, that $\om^*(X_F+X_\chi,X_G)\es\om^*(X_F,X_G)$
for any Hamiltonian vector field belonging to the constraints. Thus
$\om^*$ depends only on the tangent components of $X_F,X_G$. But
for tangent $X_F$ we have $\om(X_F,X_\al)\approx 0$ (see \ref{5.11}) and
$\om^*$ can be replaced by $\om$ without changing the value of \ref{5.14}.
This proves that $\om^*$ is just the pull-back of $\om$.
\section{First class constraints and gauge transformations}
For purely FC systems the time evolution is governed
by
\begin{equation}
\dot F\approx \{F,H\}+\{F,\phi_a\}\mu^a,
\end{equation}
where the $\phi_a$ are the primary FCC.
For the same initial conditions we get different evolutions,
depending on the multipliers $\mu^a$. The presence of arbitrary
functions $\mu^a$ in the primary Hamiltonian tells us that not
all $x\es(q,p)$ are observable, i.e. there are several $x$
representing a given physical state.
Assume that the initial value $x(0)$ is given and represents
a certain state. Then the equation of motion should fully determine
the physical state at later times. So if
$x^\pr(t)\neq x(t)$ stem from the same physical state
$x(0)$ they should be identified.\par
Consider two infinitesimal time evolutions of $F=F(0)$ given
by $H_p$ with different values of the multipliers,
\begin{equation}
F_i(t)=t\{F,H\}+t\{F,\phi_a\}\mu^a_i\mtxt{i=1,2}.\label{5.16}
\end{equation}
The difference $\delta F\es F_2(t)\ms F_1(t)$ between the values is then
\begin{equation}
\delta_\mu F=\{F,\mu^a\phi_a\},\mtxt{,}\mu=t(\mu^2-\mu^1).
\end{equation}
Such a transformation does not alter the the physical state at time $t$,
and hence is called {\it gauge transformation} \cite{d64}. Now we calculate
\begin{equation}
\big[\delta_\mu,\delta_\nu\big] F=\{\{\mu^a\phi_a,\nu^b\phi_b\},F\}
\end{equation}
and conclude that the commutator of any two primary FCC also generate
gauge transformations.
Also, performing a gauge transformation at $t\es 0$
with multipliers $\nu$ and then time evolve with multipliers $\mu$
should lead to the same state as doing these transformations in
the reverse order. We find
\begin{equation}
\big[\delta_{t,\mu},\delta_\nu\big] F
=-t\{\{\nu^a\phi_a,H\},F\}-t\{\{\nu^a\phi_a,\mu^b\phi_b\},F\}
\end{equation}
and conclude that the commutators $\{\phi_a,H\}$ also generates
gauge transformations.\par
We have seen that the first class functions
are closed with respect to the Poisson bracket and thus
the $\{\phi_a,\phi_b\}$ and $\{\phi_a,H\}$ are linear
combinations of the FCC. However, in general there will
appear secondary FCC in these combinations. Also, if we
compared the higher order terms in the time evolutions
\ref{5.16} we would find that time derivatives of $\{\phi_a,H_p\}$
generate gauge transformations. This way secondary FCC show up
as gauge transformations in all relevant systems. Therefore,
Dirac conjectured that
{\it all} FCC $\gam_a$ generate gauge transformations. We shall
assume this conjecture to hold in what follows although
there are some exotic counterexamples \cite{a75,c79}.\par
Note, however, that if the structure constants in
\begin{equation}
\{\gam_a,\gam_b\}=t_{ab}^{\;\,c}\gam_c\label{5.18}
\end{equation}
depend on the canonical variables, then $[\delta_\mu,\delta_\nu]F$
is a gauge transformation only on the constraint surface. Also, above we
made the hidden assumption that time (or the space-time coordinates in field
theory) is not affected by the transformation. Else we would have
to take $F+\delta_\mu F$ at the transformed time $t+\delta_\mu t$
before calculating the second variation $\delta_\nu$.
We come back to this point when discussing generally covariant
theories in chap.7.\pan

We conclude that the most general physically permissible motion
should allow for an arbitrary gauge transformation to be
performed during the time evolution. But $H_p$ contains
only the primary FCC. We thus have to add to $H_p$ the secondary FCC
multiplied by arbitrary functions.
This led Dirac to introduce the {\it extended Hamiltonian}
\begin{equation}
H_p\longrightarrow H_e=H+\cn^a\gam_a
\end{equation}
which contains {\it all}  FCC \cite{d64}. $H_p$ accounts for all the gauge
freedom.\par
Clearly, $H_p$ and $H_e$ should imply the same time
evolution for the classical observables.
{\it Observables} are gauge invariant functions on $\Gamma_c$,
that is phase space functions that weakly commute with
the gauge generators,
\begin{equation}
F\mtxt{observable}\Longleftrightarrow \{F,\gam_a\}\approx 0\mtxt{for all}a.
\end{equation}
Since $H_e-H_p$ is a combination of the secondary FCC, we have
\begin{equation}
\dot F\approx\{F,H_p\}\approx \{F,H_e\}
\end{equation}
for any observable $F$, as required. In the extended formalism
one makes no distinction between primary and secondary FCC since they
are treated symmetrically. The introduction of $H_e$ is a new
feature of the Hamiltonian scheme. It does not follow from
the Lagrangean formalism.
\par
Let us now investigate the {\it geometric meaning} of FC systems.
As a preparation we show:\vskip .3truecm\pan
The induced 2-form $j^*\om$ has rank $\geq N-2M$,
where $M$ is the number of independent first or second class
constraints.\vskip .3truecm\pan
Let us assume that the tangent vectors $X_p,\,p\es 1,\dots,P$ form a basis
for the null-eigenvectors of $j^*\om$, i.e. $j^*\om(X_p,Y)\es
\om(X_p,Y)=0$ for all tangent $Y$. Let us now see how big $P$ can
be. For that we consider
\begin{equation}
a^p\om(X_p,Z_q)=0,\mtxt{where the} Z_q,\quad q=1,\dots,M
\end{equation}
together with the tangent vectors form a basis of $T\Gamma$
at the point under consideration. These are $M$ equations
for $P$ unknown. So, if $P\geq M$ then there would always
exist a solution $X\es a^pX_p$ with $\om(X,Z_q)=0$ for all $Z_q$.
Being also a null-eigenvector of $j^*\om$ we would conclude
that $\om(X,Z)\es 0$ for all vectors $Z\in T\Gamma$ or that $\om$ is
degenerate. This then proves the statement above. Now we have the
following
\begin{satz}
For a FC system the induced two-form $j^*\om$
is maximally degenerate. The kernel is spanned
by the Hamiltonian vector fields belonging to the FCC.
\end{satz}
First, if $X^\mu\pa_\mu\gam_b\approx 0$ for all constraints, then
$X$ is tangent. But since $X^\mu_a\pa_\mu\gam_a\approx\{\gam_a,
\gam_b\}\approx 0$, all Hamiltonian vectorfields belonging to
the constraints are tangent. Second, for an arbitrary tangent vector $X$
we have $\om(X_a,X)\es i_{X_a}(X)\es \pa_\mu\chi_a X^\mu\approx 0$.
Thus the $M$ $X_a$'s are null-eigenvectors of the induced
two-form and the rank of $j^*\om$ equals $2N\ms 2M$, i.e.
it is maximally degenerate.\par
Thus we have the following situation: The $M$ FCC
generate {\it flows which stay on the constraint surface}
and which we identified with gauge transformations.
The Hamiltonian vector fields belonging to the constraints
are the null-directions of the induced pre-symplectic $2$-form.
That these null-vector fields generate gauge orbits follows from
\begin{satz} On $\Gamma_c$ the vectors $X_a$ generate $M$-dimensional
manifolds.\end{satz}
The proof uses the Frobenius
integrability condition, which says that $M$ linearly independent
vector fields are integrable (through each point in $\Gamma_c$ there is
a surface, the gauge orbit, to wich the $X_a$ are tangent)
iff all Lie brackets $[X_a,X_b]$ are linear
combinations of $(X_1,\dots,X_M)$. Indeed,
\begin{equation}
[X_a,X_b]=-X_{\{\gam_a,\gam_b\}}=
-t_{ab}^{\;c}X_c+\gam_c\om^{\mu\nu}\pa_\nu t_{ab}^{\;c}
\approx -t_{ab}^{\;c}X_c,
\end{equation}
where we used \ref{5.18}. Note that for $(q,p)$-dependent structure
constant (as in gravity) the null vector fields are integrable
only on the constraint surface.

In a next step one wants to eliminate the gauge degrees of freedom
that is indentify points on the same gauge orbit.
This can in principle be achieved by introducing
gauge invariant variables, e.g. the
transverse potential or holonomies in electrodynamics, or alternatively
by fixing the gauge. A {\it gauge fixing} must obey two conditions: first
it must be attainable and second it should fix the gauge uniquely
\footnote{There may be obstructions to fulfilling these requirements as
has been demonstrated by Gribov and Singer \cite{g78,s78}.}.
We can fix the gauge by imposing the independent conditions
\begin{equation}
F_a(q,p)=0,\mtxt{a=1,\dots,M.},
\end{equation}
The surface defined by these conditions should intersect every
gauge orbit in exactly one point. A necessary condition
is that at least one gauge fixing function $F_b$ should
change in the direction of all null-vectors $X_a$. In other words, there
is at least one $F_b$ so that
\begin{equation}
\lam^a(X_a,\nabla F_b)=\lam^a\{\gam_a,F_b\}\neq 0
\end{equation}
for all $\lam$. This implies that
\begin{equation}
\det\{\gam_a,F_b\}\equiv\det\, F_{ab}\neq 0.\label{5.24}
\end{equation}
In particular, if we could choose gauge fixings canonically conjugated
to the constraints, $\{\gam_a,F_b\}=\delta_{ab}$, then the
gauge orbits would intersect the gauge fixing surfaces orthogonal
and in this case $\det F_{ab}=1$.
The determinant of $F$ plays an important part in the
quantization of gauge systems and is the wellknown
{\it Fadeev-Popov determinant} \cite{fp67}.\par
Because of \ref{5.24} the FCC together with the gauge fixings form a SC system
and we can take over the result from the previous subsection.
The reduced phase space $\Gamma_r$ consists of the points fulfilling the
constraints and gauge fixings. Collecting the $\gam_a$ and $F_a$ into one
vector, $\Omega_p,\,p=1,\dots,2M$, we find for
the Hamiltonian equation of motion for any phase space
function
\begin{equation}
\dot F=\{F,H\}-\{F,\Omega_p\}G^{pq}\{\Omega_q,H\}.
\end{equation}
\section{Mixed second and first class constraints}
Before gauge fixing the evolution is governed by the
first class partner of the extended Hamiltonian
\begin{equation}
H_e^*=H+\gam_a\cn^a-\chi_\al\Delta^{\al\beta}\{\chi_\beta,H\}
\end{equation}
since we have
\begin{equation}
\dot F=\{F,H_e^*\}=\{F,H\}+\{F,\gam_a\}\cn^a-\{F,\chi_\al\}\Delta^{\al\beta}
\{\chi_\beta,H\}.
\end{equation}
For a discussion of the {\it starred variables} see \cite{bk60}.
After gauge fixing one can again introduce starred variables
with respect to the SC system $\psi_I=(\chi_\al,\gam_a,F_b)$.
Denoting the Poisson brackets matrix of
all these constraints $\psi_I$ by $\tilde\Delta$, we have
\begin{equation}
\dot F=\{F,H^*\},\mtxt{where}H^*=H+\psi_I\tilde\Delta^{IJ}
\{\psi_J,H\}.
\end{equation}
\section{Gauge fixing of Chern-Simons theory}
In section $4$ we have seen that the Chern-Simons Lagrangean \ref{4.1}
leads to two SCC and two FCC. Now
we supplement those by two gauge fixing conditions, namely
\begin{equation}
F_1=A_0\mtxt{and}F_2=\pa_i A^i.
\end{equation}
Altogether the conditions $(\chi_1,\chi_2,\gam_1,F_1,\gam_2,F_2)$
form a SC system  and define $\Gamma_r$.
The surface defined by $\chi_i, \gam_1$ and $F_1$ can be parametrized
by the spatial components of the gauge potential which can be
decomposed as
\begin{equation}
A_i=\eps_{ij}\pa_j\varphi+\pa_i\lambda+{1\ov L}q_i\label{5.30a}
\end{equation}
with constant $q_i$ \footnote{The $U(1)$-bundle over the torus
is non-trivial and $A$ must be periodic only up to non-trivial
gauge transformations \cite{sw92}. For simplicity we assume here
that $A$ is periodic and hence $\int F_{12}\es 0$}.
Imposing further $\gam_2$ and $F_2$ we see that
\begin{equation}
A_i=-{1\ov\kappa}\eps_{ij}\pa_j{1\ov \lap}J^0+{1\ov L}q_i\label{5.30b}
\end{equation}
so that $\Gamma_r\es \{q_1,q_2\}$ is finite-dimensional.
Furthermore, $\gam_2\es 0$ and the periodicity of the $A_i$
imply that the total charge $Q\es\int d\vx J^0$ must vanish.
\par
The inverse Poisson bracket 'matrix' reads
\begin{equation}
(\tilde\Delta^{IJ})(x,y)={1\ov\kappa}\pmatrix{0&-1&0&0&{1\ov\lap}\pa_2&0\cr
                  1&0&0&0&-{1\ov\lap}\pa_1&0\cr
                  0&0&0&\kappa&0&0\cr
                  0&0&-\kappa&0&0&0\cr
                  {1\ov\lap}\pa_2&-{1\ov\lap}\pa_1&0&0&0&-\kappa{1\ov\lap}\cr
                  0&0&0&0&\kappa{1\ov\lap}&0\cr}\delta(x-y)
\end{equation}
and one finds the following Dirac bracket for the coordinates
on $\Gamma_r$
\begin{equation}
\{q_i,q_j\}=-\delta_{ij}.
\end{equation}
Calculating the starred Hamiltonian, one needs to remember that
for a periodic function one has
$\lap^{-1}\lap f=f-V^{-1}\!\int f$. After some algebra one finds
\begin{equation}
H^*=-\int d\vx \Big\{A^0\big(\kappa F_{12}-J_0\big)+{1\ov \kappa}
J_0{1\ov\lap}\eps_{ij}\pa_i J_j\Big\}
-\ha j_i q^i-{1\ov\kappa}\eps_{ij}j_ip_j,
\end{equation}
where we have introduced the mean 'fluxed'
\begin{equation}
j_i\equiv {1\ov L}\int d\vx \,J_i\mtxt{,}
q^i\equiv {1\ov L}\int d\vx \,A^i\mtxt{and}
p_i\equiv {1\ov L}\int d\vx \,\pi_i.
\end{equation}
After imposing the constraints $\chi_i,\gam_2$ and $F_1$ the
non-trivial equations of motion take the simple form
\begin{equation}
\dot q_i=-{1\ov\kappa}\eps_{ij}j_j.\label{5.32}
\end{equation}
Of course, the evolution belonging to $H^*$ stays on $\Gamma_r$.
Since $\Gamma_r$ is $2$-dimensional the (topological) Chern-Simons
theory is effectively a simple mechanical system. This was expected
from the beginning since there are $6$ constraints and gauge fixings
for $6$ degrees of freedom (per $\vx$).
\par
To see the meaning of this result more clearly, let us see
what the observables are. As coordinates on $\Gamma_c$
we may take $\lam$ and $q_i$ in \ref{5.30a}, so that we are
considering functionals $F[A^0,\lam,q_i,J_a]$.
Such $F$ commute with the FCC if they
are independent of $\lam$ and $A^0$. Hence observables
have the form
\begin{equation}
F=F[J_a,q_i]
\end{equation}
and depend only on the zero-modes of the $A_i$.\par
Let us finally remark that for a pure CS theory ($J\es 0$)
the Lagrangean density is invariant, up to a total time
derivative, under global gauge transformations for which
only $e^{i\lam}$ must be periodic. This
introduces global gauge transformations with windings
around the handels of the torus (the box with opposite points
identified). Hence we must identify gauge potentials
which are related by such global gauge transformations
\begin{equation}
A_i\sim A_i+{2\pi\ov L}n_i\mtxt{or}
q_i\sim q_i+2\pi n_i .
\end{equation}
Gauge invariant functionals must be invariant under such
transformations. Thus they depend only on $\exp(i\sum m_iq_i)$.
For pure Chern-Simons theories we have
\begin{equation}
e^{i\sum m_iq_i}=W(C,A)=\exp\Big\{i\oint_C A\Big\}
\end{equation}
on the constraint surface ($F_{12}\es 0$) if the loop $C$ winds
$m_i$-times around the torus in the direction $i$. For a contractible
loop $W(C,A)$ vanishes. Thus, observables have the
form
\begin{equation}
F(A)=F\Big(e^{i\oint_C A}\Big).
\end{equation}
Let $C,D$ be $2$ loops which wind $m_i,n_i$-times around the
torus in the direction $i$. We parametrize them by $x(\tau),y(s)$.
We compute
\begin{equation}
\oint\limits_{C}\oint\limits_{D}\{A(x(\tau)),A(y(s))\}=
-\oint\limits_{C}\oint\limits_{D}\eps_{ij}\dot x^i(\tau)
\dot y^j(s)d\tau ds=-(n_1 m_2-n_2 m_1).
\end{equation}
Upon deformation of the curves the commutator is invariant and
therefore is a {\it topological invariant}. This can be understood by noting
that for $J\es 0$ the Chern-Simons
model \ref{4.1} is invariant under space-time diffeomorphisms.
In particular the {\it spatial} ones are generated by
\begin{equation}
G_{diff}=\int d\vx \eps^iA_i\gam_2
\end{equation}
and hence observables must be invariant under spatial diffeomorphisms.
\par
Finally note, that for the pure Chern-Simons theory the phase
space variables $q_i$ lie in $[0,2\pi]$, that is $\Gamma_r$ is compact
and as a consequence the Hilberspace becomes finite dimensional.

\section{First order action principle and symmetries}
The solutions to the primary Hamiltonian equations of motion \ref{3.12a},
\ref{3.12b} extremize the primary (or total) first order action,
\begin{equation}
\delta S_p=\delta\int\limits_{t_1}^{t_2}\Big(\dot\qi\npi-H-
\sum_{primary}u^m\phm\Big)dt=0\label{5.35}
\end{equation}
with respect to variations $\delta q,\delta p,\delta u$ subject
only to the restriction $\delta q(t_1)\es\delta q(t_2)\es 0$. The
variables $u^m$ which have been introduced to make the Legendre
transformation invertible appear now as Lagrange multipliers
enforcing the {\it primary constraints}. It is clear that the theory
is invariant under $H\to c^m\phm$ since such a change can be
absorbed into the Lagrange multipliers.\par
The variational principle \ref{5.35} is equivalent to
\begin{equation}
\delta \int \Big(\dot\qi\npi-H\Big)dt=0\mtxt{subject to
}\phm=\delta\phm=0.
\end{equation}
There is yet another variational principle which for gauge invariant
observables leads to the same time evolution, namely the {\it extended
action principle}. The equations of motion for the extended
formalism follow from
\begin{equation}
\fbox{$\displaystyle
\delta S_e=\delta\int\Big(\dot\qi\npi-H-\sum_{all\,constr.}u^j\phi_j\Big)dt
 =0, $}\label{5.37}
\end{equation}
where the sum extends over {\it all constraints}.\par
Take the case of purely SCC and
let $y^i\to x^\mu(y^p)$ be the imbedding of $\Gamma_r\subset \Gamma$.
The Lagrange multiplier method guarantees that the implementation of the
constraints $\chi_\al$, either directly or via the Lagrange
multipliers, are equivalent. Now let us solve the constraint
directly in \ref{5.37}. Recall that a symplectic
$2$-form can locally be written as $\om=d\theta$. The pull-back
of the one-form potential $\theta$ is
\begin{equation}
j^*(\theta)=j^*(\npi d\qi)\equiv j^*(a_\mu dx^\mu)=
a_\mu(x(y)){\pa x^\mu \ov \pa y^p}dy^p.
\end{equation}
Inserting this into the extended action one finds
\begin{equation}
S_e[y]=\int\big\{\theta -H(x(y))dt\big\}=\int \big(a_p \dot y^p-h\big)dt.
\label{5.38}
\end{equation}
The corresponding variational principle yield then the
equation of motion for SC systems:
\begin{equation}
\delta S_e[y]=0 \Longleftrightarrow
\dot y^p=\{y^p,h(y)\}^*.
\end{equation}
This can be checked directly by using
\begin{equation}
j^*\om=j^*d\theta=d\,j^*\theta=d\,a_p(y)dy^p.\label{5.40}
\end{equation}
The fact that \ref{5.37} yields \ref{5.38} is of practical use when calculating
Dirac bracket. One solves for the constraints inside the action
and from the new kinetic term one reads off the induced potential
form on $\Gamma_c$. From \ref{5.40} one computes the induced symplectic
form and thus the Dirac bracket.\par
For FC systems it is also legitimate to solve the FCC
inside the action
\begin{equation}
S_e=\int\Big(\qi\npi-H-\cn^a\gam_a\Big)dt.\label{5.41}
\end{equation}
However, since the induced $2$ form is degenerate the equations
of motion on $\Gamma_c$ are not canonical. To get Hamiltonian
equations one needs to go to $\Gamma_r$ by
imposing additional gauge conditions. Then one may write down
the corresponding action for the SC system as discussed above.\par
Let us demonstrate how this works for the {\it Chern-Simons theory}.
Inserting the fields on the reduced phase space \ref{5.30b} into
the first order action results in
\begin{equation}
S=\int dt\int \big({\kappa\ov 2}q_i\eps_{ij}\dot q_j-q_ij_i\big)
+{1\ov \kappa}\int d\vx
J^0{1\ov\lap}\eps_{ij}\pa_iJ_j
\end{equation}
which of course reproduces the correct equations of motion \ref{5.32}.
\chapter{Yang-Mills Theories}
In this chapter I consider YM theories \cite{al73} without coupling to matter
and emphasize the role of the constraints \cite{hrt76,s82}. Pure {\it
non-abelian} YM theories are interesting in their own right and they
are highly non-trivial.\par
The YM action for the gauge fields is
\begin{equation}
S=-{1\over 4}\int \tr\big[F_{\mu\nu}F^{\mu\nu}\big]d^3x dt,
\end{equation}
where \footnote{$a,b,\dots$ denote internal indices, $\mu,\nu\dots$
space-time indices. The $T_a$ are hermitean generators and the
structure constants $\str$ are totally antisymmetric.} the field strength
is Lie-algebra valued,
\begin{equation}
F_{\mu\nu}=\pamu A_\nu-\panu A_\mu -i[A_\mu,A_\nu]\mtxt{,}
A_\mu= A^a_\mu T_a\mtxt{,}[T_a,T_c]=i\str T_c,
\end{equation}
and the action is invariant under local gauge transformations
\begin{equation}
A_\mu\longrightarrow e^{-i\theta}A_\mu e^{i\theta}+ie^{-i\theta}
\pamu e^{i\theta}
\end{equation}
with $\theta=\theta^a T_a$. The functions $\theta^a=\theta^a(x,t)$
are arbitrary functions on space-time. The infinitesimal form of
these gauge transformations is
\begin{equation}
\delta_\theta A_\mu^a=-\big(\pamu \theta^a +f^a_{b c}A_\mu^b \theta^c\big)
=-(D_\mu\theta)^a.\label{6.4}
\end{equation}
The local gauge invariance implies generalized Bianchi identities
\begin{equation}
D_\mu D_\nu F^{\mu\nu}=0
\end{equation}
and renders the system singular. Among the field equations
$D_\mu F^{\mu\nu}$ there are some containing second
time-derivatives of $A$,
\begin{equation}
D_\mu F^{\mu i}=0\mtxt{,}i=1,2,3
\end{equation}
and which therefore are dynamical equations of motion. The others
\begin{equation}
D_\mu F^{\mu 0}=D_i F^{i0}\mtxt{or}
\phm(A,\dot A)=\pa_i F_m^{i0}+f_{ab}^{\;\,m}A_i^a F_b^{i0}=0,
\end{equation}
where $m\es 1,\dots,N$=dim(Gauge Group),
are Lagrangean constraints. No further constraints appear
since the time derivatives of the $\phm$ vanish on account of
the field equations and the constraints themselves.\par
The canonical momenta conjugate to the $A$'s are
\begin{equation}
\pi_a^\mu=-F^{0\mu}_a\mtxt{,} \{A_\mu^a(\vx),\pi_b^\nu(\vy)\}
=\delta^a_b\delta_\mu^\nu\delta(\vx-\vy).
\end{equation}
Since the field strength tensor is antisymmetric we obtain
$N$ {\it primary constraints}
\begin{equation}
\phm (A,\pi)=\pi^0_m\approx 0.
\end{equation}
After a partial integration the canonical Hamiltonian is found to be
\begin{equation}
H=\int d\vx\Big(\ha\pi_i^a\pi_i^a+{1\ov 4}F_{ij}^aF_{ij}^a-
A_0^aD_i\pi^i_a\Big),
\end{equation}
and determines the time evolution
\begin{equation}
\dot F=\{F,H_p\}\mtxt{,}H_p=H+\int d\vx\, u^m\phm.
\end{equation}
We need to check the consistency of the primary constraints:
\begin{equation}
\dot\phm =\{\phm,H_p\}=0\Longrightarrow
\tilde\phm=(D_i\pi^i)_m\approx 0.
\end{equation}
These $N$ secondary constraints are the generalizations of
the {\it Gauss constraint} in electrodynamic.\par
The only non-trivial Poisson brackets of the algebra of constraints
are
\begin{equation}
\{\tilde\phm(\vx),\tilde\phi_n(\vy)\}=
f_{mn}^{\;\,p}\tilde \phi_p(\vx)\delta(\vx-\vy).
\end{equation}
The algebra is closed and therefore the $2N$ constraints
$\gam_a\es(\phm,\tilde\phi_n)$ form a FC system.
Their Poisson brackets with $H$ are computed to be
\begin{equation}
\{\phm,H\}=\tilde\phm\approx 0\mtxt{,}
\{\tilde\phm ,H\}=-f_{mn}^{\;\,p}A_0^n\tilde\phi_p\approx 0.
\end{equation}
Let us now investigate the relation between the Hamiltonian gauge
symmetries generated by the FCC and the
Lagrangean gauge transformations \ref{6.4}. A general
combination of the FCC $\phi=\int(\eps_1^m\phm+\eps_2^m\tilde\phm)$
generates the canonical symmetries
\eqngr{\delta A_\mu^a&=&\{\phi,A_\mu^a\}=\delta_\mu^0\eps_1^a-
\delta_\mu^i D_i\eps_2^a}{
\delta \pi^\mu_a&=&\{\phi,\pi^\mu_a\}=\delta^\mu_i f_{ab}^{\;\,c}
\eps_2^b\pi^i_c+\int\phm\{\eps_1^m,\pi_a^m\},}
where we have already anticipated that $\eps_1$ depends on $A_0$.
{}From \ref{6.4} we read off how the $\eps$'s must be chosen to
correspond to Lagrangean gauge transformations. We find
that the particular combination
\begin{equation}
G=D_0\theta^m\phm-\theta^m\tilde\phm
\end{equation}
generates those transformations. Both primary and secondary
FCC enter the Lagrangean gauge transformations similarly as
for the CS theory.\par
Alternatively we can introduce gauge invariant variables, e.g.
the Wilson loops \cite{wy75}, (see the lecture of R. Loll)
or fix the gauge. To fix the gauge freedom we
need $2N$ gauge fixing conditions
on the phase space variables $(A,\pi)$. Contrary to the
situation in electrodynamics the gauge fixing
in YM theories is rather subtle due to the
Gribov problem \cite{g78}. Let $F_a(A_\mu)$ be local gauge fixings (which
we assume not to depend on the momenta). Then the following
problem may arise:\pan
{\it There are several field $A_\mu^{(j)}$ which are related
by finite gauge transformations and all of them obey the
gauge fixing.}  \vskip 0.1truecm\par
This happens for example for $QED_2$ on the euclidean torus, if
one decomposes a gauge field as in \ref{5.30a}.
The local condition $\pa^\mu A_\mu$ eliminates the gauge function
$\lam$ but does not constrain the $q_i$. But $2\pi$ and
$q_i\!+\!2\pi$ are gauge equivalent configuration and
this freedom cannot be fixed by a local gauge conditions.
This is an example to a more general situation which
has been proven by Singer \cite{s78}: For compactified YM-theories
no global continuous gauge choice of the (local) form $F_a(A)\es 0$
exists which completely specifies the gauge.\par
Rather then dwelling on the various gauge fixings, their
merits and drawbacks, which is of course important in
a functional quantization, let me make some remarks about
the variational problem.\par
The primary FCC $\phm$ are sort of uninteresting,
since they can be easily eliminated in the extended formalism.
We eliminate them by setting $\pi^0\es A_0\es0$ and find the first
order action
\begin{equation}
S=\int \Big[\pi_a^i\dot A^a_i-
\cn^a\gam_a-\ha (\pi_a^i\pi^a_i+B^i_a B^a_i)\Big]dt\mtxt{,}
\gam_a=(D_i\pi^i)^a,
\end{equation}
with multiplier fields $\cn_a$.
This form of the action is the one which is usually met in the literature
(for example, in gravity one does not keep
the momenta conjugated to the lapse and shift functions in
the first order action). After having eliminated one pair
of canonical variables one may wonder how one can reconstruct
the full set of Lagrangean gauge transformation \ref{6.4}.
This is indeed possible for all relevant systems in physics,
e.g. YM-theories, the relativistic particle, string and
gravity \cite{mw92}. We will come back to this point in the next chapter.
\chapter{Lagrangean Symmetries of First-Class Hamiltonian Systems}
\def\ta{{\tilde a}}
\def\ti{{\,\tilde i}}
\def\tj{{\,\tilde j}}
\def\hr{\hat{R}}
\def\hi{\hat{I}}
\def\jm{\hat{N}}
\def\hio{\hat{I}_{\xi_1\lambda_1}}
\def\hit{\hat{I}_{\xi_2\lambda_2}}
\def\dxila{\delta_{\xi,\lambda}}
\def\tttr{t^\tga_{\tal \tbe}}
\def\tttw{t^\tbe_\tal}

\section{The relativistic particle}
We describe the relativistic particle
moving in $4$-dimensional Minkowski spacetime by $4$ scalar fields
$\phi^\mu(t)$, $\mu=0,1,2,3$, in $1$-dimensional 'spacetime'.
The action for the relativstic particle takes the form
\begin{equation}
S=-\ha\int\sqrt{-g}\big[g^{00}\dot\phi^\mu\dot\phi_\mu+m^2\big]dt\label{7.1}
\end{equation}
where the overdot denotes differentiation with respect to time $t$ and
$\phi^\mu\phi_\mu=-(\phi^0)^2+\sum_1^3 (\phi^i)^2$. The
$m^2$-term maybe viewed as 'cosmological constant' in $1$-dimensional
'spacetime'.

$S$ is invariant with respect to general
coordinate transformations (repara-metrization invariance).
The infinitesimal form of these transformations reads
\begin{equation}
t\to t-\xi,\qquad g_{00}\to g_{00}+\cl_\xi g_{00},\qquad
\phi^\mu\to \phi^\mu+\cl_\xi \phi^\mu,\label{7.2}
\end{equation}
where $\cl_\xi$ is the Lie-derivative.
Introducing  the lapse function $\cn$ according to
\begin{equation}
g_{00}=-\cn^2
\end{equation}
we get the following transformation law from \ref{7.2}
\begin{equation}
\delta\phi^\mu=\dot\phi^\mu\xi\quad\hbox{and}\quad
\delta \cn=(\cn\xi)^\cdot\;.\label{7.4}
\end{equation}
The action \ref{7.1} leads to the primary constraint
$\phi_1\es \pi_{g^{00}}\es 0$ which in turn implies
the secondary one
\begin{equation}
\phi_2\equiv \gam=\ha\big(\pi^\mu \pi_\mu +m^2\big).
\end{equation}
These are FCC. The partial
gauge fixing $F_1\es g^{00}\!+\!1\es 0$ and $\phi_1$
form a conjugate second class pair
and can be eliminated. Applying the standard procedure
one finds then the following first order action
\begin{equation}
S=\int\big[\pi_\mu\dot\phi^\mu-\cn \gam\big]dt,\label{7.6}
\end{equation}
where the $\pi_\mu$ are the momenta conjugated to the $\phi^\mu$,
\begin{equation}
\{\phi^\mu,\pi_\nu\}=\delta^\mu_\nu.
\end{equation}
The lagrangean multiplier $\cn$ accompanying the constraint
$\gam$ (which is the super-hamiltonian) reintroduces the
lapse function.\par

The action \ref{7.6} is invariant with respect to the infinitesimal
off mass-shell {\it gauge} transformations
\begin{equation}
\delta_\lam \phi^\mu=\{\phi^\mu,\lam \gam\}
=\pi^\mu\lam,\quad
\delta_\lam \pi_\mu=\{\pi_\mu,\lam \gam\}=0,\quad
\delta_\cn=\dot{\lam}.\label{7.8}
\end{equation}
With the identification $\lam=\cn\xi$ these transformations coincide
with the diffeomorphism transformations \ref{7.4}), but {\it only
on mass shell}:
\begin{equation}
\dot\phi^\mu=\cn \pi^\mu,\qquad \dot \pi^\mu=0.
\end{equation}
This is a general problem with diffeormorphism invariant
theories. Only on-shell can the transformations generated by the FCC
alone be identified with the Lagrangean symmetries.
On the technical side the difficulty
of identifying gauge and diffeomorphism transformations
can be traced back to the nonlinear dependence
of the constraint on the momentum. This is the important
difference between internal and spacetime symmetries.
In the following section we shall
see how the canonical transformations generated by the FCC
must be modified to yield all Lagrangean symmetries.
\section{Hamiltonian vs. Lagrangean symmetries}
In this chapter I shall consider a general FC system, the first order
action of which is given by \ref{5.41}. These actions
describes both systems with a finite
or infinite number of degrees of freedom if the following
condensed notation \cite{dw65} is assumed:
For systems with a finite number of degrees of freedom
$a$ and $i$ are discrete and for field theories they denote
both internal indices and space-coordinates. To distinguish internal
from composite indices we shall use tildes for the latter ones.
For field theories $\ti =\{i,\vx\} $ and $\tal =\{a,\vx\}$,
where $i$ and $a$ are some discrete (internal) indices.
For a scalar field $q^\ti(t)\es \varphi^x(t)\es \vph(x,t)$
and for a vector field $q^\ti(t)=A^{i,x}(t)=A^i(x,t)$.
We adopt the Einstein convention and assume summation over
discrete repeated indices and integration over continuous ones, for
example
\begin{equation}
\xi^x p_{i,x}\dq^{i,x}=\sum_i\int dx \xi(x)p_i(x)\dq^i(x)
\end{equation}
but
\begin{equation}
\xi^x p_i^x q^{ix}=\sum_i \xi(x)p_i(x)q^i(x),\qquad\hbox{no integration.}
\end{equation}
Also, we shall not distinguish $q^{i,x}$ and $q^i_x$ and use the
position of the continuous index just to indicate when we should
integrate. Sometimes it will be convenient to resolve the
composite index $\ti$ (or $\tal$) as $i,x$ (or $a ,x$).
If the system contains fermions then
some of the variables $p,q,\cn$ will be of Grassmannian type.\par
In particular the first order action reads
\begin{equation}
S_e=\int \Big(p_\ti\,\dq^\ti-\cn^\tal\gam_\tal-H\Big)dt.
\label{7.10}
\end{equation}
For FC systems the contraints
and Hamiltonian form a closed algebra
(possibly extended to fermionic variables, in which case the algebra
is graded \cite{c76}):
\begin{equation}
\{\gam_\tal ,\gam_\tbe\}=\tttr \gam_\tga\qquad \hbox{and}\qquad
\{H,\gam_\tal\}=\tttw \gam_\tbe,
\end{equation}
with structure coefficients  $\tttr$ and $\tttw$. These
coefficients may depend on the canonical variables $p,q$. \par
The equation of motion resulting from the variation of the
action \ref{7.10} with respect to $q,p$ and the Lagrangean multipliers
$\cn$ are
\begin{equation}
\delta S_e=\int\Big(\delta p_\ti EM(q^\ti)-\delta q^\ti EM(p_\ti)
-\delta \cn^\tal \gam_\tal\Big) dt +\hbox{bound. terms}
\end{equation}
are
\begin{eqnarray}
EM(q^\ti)&\equiv& \dot{q}^\ti-\{q^\ti,\cn^\tbe \gam_\tbe+H\}=0,\nonumber\\
EM(p_\ti)&\equiv& \dot{p}_\ti-\{p_\ti,\cn^\tbe \gam_\tbe+H\}=0,\\
\gam_\tal&=&0.\nonumber
\end{eqnarray}
We use the abbreviations $EM(q)$ and $EM(p)$
for the equations of motion. Of course, on mass shell we have
$EM\es 0$, but off mass shell either $EM(q)$ or $EM(p)$ (or both)
does not vanish.

The first order action is invariant with respect to the infinitesimal
transformations \cite{mw92}
\begin{equation}
\hi_{\xi,\lam}\,F(q,p,\cn )=F\big(\hi_{\xi,\lam}\;q,
\hi_{\xi,\lam}\;p,\hi_{\xi,\lam}\;\cn\big),\qquad
\hi_{\xi,\lam}=\hat{1}+\dxila+\cdots,\label{7.14a}
\end{equation}
where
\begin{eqnarray}
\dxila\, q^{ix}&=&EM(q^{ix})\xi^x + \{q^{ix},\lam^\tbe \gam_\tbe\},\nonumber\\
\dxila\, p_i^x&=&EM(p_i^x)\xi^x+\{p_i^x,\lam^\tbe \gam_\tbe\},\label{7.14b}\\
\dxila\, \cn^\tal &=&\dot{\lam}^\tal-\lam^\tbe\cn^\tga t^\tal_{\tga\tbe}
-\lam^\tbe t^\tal_\tbe.\nonumber
\end{eqnarray}
The parameters $\xi^x=\xi(x,t)$ and $\lam^\tal=\lam^a
(\cn,x,t)$ are the parameters of the infinitesimal transformations.
The first order Lagrangean is invariant
up to a total time-derivative and correspondingly the action up to
boundary terms
\begin{equation}
\dxila S_e=\Big(p_\ti\dxila \;q^\ti-\lam^\tal \gam_\tal\Big)\vert_{t_i}^{t_f}.
\end{equation}
If the parameters $\xi,\lam^\tal$ vanish at the initial and final times
then $\dxila S=0$. If we would like the action to be invariant
even under transformations with non-zero $\xi,\lam$ at the boundaries
then we need to add to the action the total derivative of some
function $Q(q,p)$ which satisfies the equation
\begin{equation}
{\delta Q\over \delta q^\ti}\dxila q^\ti +
{\delta Q\over \delta p_\ti}\dxila p_\ti =\dot{\lam}^\tal \gam_\tal-
p_\ti\dxila q^\ti,
\end{equation}
where $\delta/\delta q$ etc. means functional derivative with respect
to $q$.

To construct the finite transformations we need to apply
the infinitesimal transformations many times. To be successful
in this 'exponentiation' it is clear that the following
necessary condition should be fulfilled: The algebra of infinitesimal
transformations should be closed. To check the algebra of
transformations let us calculate the result for the commutator of two
subsequent infinitesimal transformations \ref{7.14a},\ref{7.14b}
with parameters $\xi_1,\lam_1$ and $\xi_2,\lam_2$ respectively.
If we are making two subsequent infinitesimal transformations,
then $\lam^\tbe$ of the second transformation will depend on $q,p$
if the structure constants depend on the canonical variables
(since $\lam^\tbe$ depends on $\cn$) and
we should keep $\lam^\tbe$ inside the Poisson brackets
even for purely bosonic theories.\par
For an arbitrary algebraic function $F(q,p)$ of the canonical
variables (for example $F\es q$ or $F\es p$) a rather lengthy but
straightforward calculation yields the commutator
\begin{eqnarray}
&\big[\hit ,\hio\big]\,F^x(q,p)=
\Big({\delta F^x\over \delta q_i^z}EM(q_i^z)+(q\to p)\Big)
(\dot{\xi}_1^z\xi^z_2-\xi_1^z\dot{\xi}_2^z)\nonumber\\
&\quad +\Big((\xi_2^x-\xi_2^y)\lam_1^\tga -
\cn^\tga \xi_2^x\xi_1^y -(1\leftrightarrow 2)\Big)
\Big(\big\{F^x,{\delta \gam_\tga\over \delta q^y_j}\big\}EM(q_j^y)+(q\to p)
\Big)\nonumber\\
&-(\xi_2^x\xi_1^y-\xi_1^x\xi_2^y)\Big(\big\{F^x,{\delta H\over \delta q_j^y}
\big\}EM(q_j^y)+(q\to p)\Big)+\{F^x,\bar{\lam}^\tga \gam_\tga\}\label{7.17a}
\end{eqnarray}
and correspondingly for the Lagrangean multipliers one has
\eqngrl{
\big[\hit,\hio]\cn^\tal&=&
\big(\hi_{\bar{\lam}}-1\big)\cn^\tal+\lam_2^\tde\lam_1^\tga\Big(
\dot{t}^\tal_{\tga\tde}-\{t^\tal_{\tga\tde},\cn^\tsi \gam_\tsi +H\}\Big)\hfill
}{
-(\lam_2^\tga\xi_1^x&-&\lam_1^\tga\xi_2^x)\Big(
{\delta\over \delta q_i^x}(\cn^\tbe t^\tal_{\tbe\tga}+t^\tal_\tga)EM(q^x_i)+
(q\to p)\Big),\hfill}{7.17b}
where we have introduced
\begin{equation}
\bar{\lam}^\tal=\lam_1^\tsi\lam_2^\tbe t^\tal_{\tbe\tsi}
+{\delta \lam_2^\tal\over \delta \cn^\tbe}\delta_{\lam_1}\cn^\tbe
-{\delta \lam_1^\tal\over \delta \cn^\tbe}\delta_{\lam_2}\cn^\tbe.\label{7.17c}
\end{equation}
We would like to stress that when we are performing the second
transformation in \ref{7.17a} \ref{7.17b} which follows the first one, then we
must use the transformed variables. In particular, instead of
$\lam_2(\cn,x,t)$ we must take $\lam_2(\hi_{\lam_1}\cn,x,t)$.
This explains the appearence of the last terms in \ref{7.17c}

In the particular case where the structure coefficents $t^\tal_{\tbe\tga}$
do not depend on the canonical variables $q,p$ the parameter $\bar{\lam}$
also does not depend on them as can be seen from \ref{7.17c}.
Also, $\dot{t}^\tal_{\tbe\tga}\es 0$ in this case and thus the commutator of
two transformations generated by the constraints only ($\xi=0$)
yields again a transformation generated by the constraint.
Hence, {\it if the structure coefficients do not
depend on the canonical variables the transformations generated
by the constraints form a closed algebra off mass-shell}.
On the other hand, if the structure coefficients do depend on the canonical
variables that does not automatically imply that the algebra of
transformations will not close. Actually,
the $q,p$-dependence
in the formula \ref{7.17c} for $\bar{\lam}$ can, in principle, be
cancelled against an appropriate choice of the $\cn$-dependence
of $\lam$. Actually this takes place
for gravity, where some of the structure coefficients depend
on $q$ \cite{mw92}. Also the last terms in \ref{7.17b} vanish in this case
on the hypersurface
\begin{equation}
\cm:EM(q)=\dot{q}^\ti -\{q^\ti,\cn^\tbe \gam_\tbe+H\}=0\label{7.18}
\end{equation}
and the algebra of transformations generated only by the
constraints is closed
on this hypersurface where the Lagrangean system lives. Although it
is still closed off mass shell, it is not for all trajectories in phase
space.\par
The algebra of transformations \ref{7.17a},\ref{7.17b} can also be closed
in all relevant cases even when $\xi\neq 0$ if the $\lam^\tbe$
and $\xi$ are related in a certain way. The corresponding
transformations are actually the symmetry transformations
corresponding to the symmetries of the Lagrangean system when
some of the constraints are nonlinear in the momenta.

Now we shall study the question when the symmetry
transformations of a Hamiltonian system are also symmetry
transformations of the corresponding Lagrangean system. For that
the transformations \ref{7.14b} should at least leave the
hypersurface $\cm$ (see \ref{7.18}) on which the Lagrangean
system lives, invariant. That is, they should leave any
trajectory which belongs to the hypersurface $\cm$
on this hypersurface. The necessary conditions for this can
be gotten by varying \ref{7.18} as follows
\eqngrl{\big(\dxila q^\ti\big) ^{\cdot} &=&
{\delta^2(H+\cn^\tsi \gam_\tsi)\over \delta p_\ti\delta q^\tj}\;\dxila q^\tj
}{&+&{\delta^2(H+\cn^\tsi \gam_\tsi)\over \delta p_\ti\delta p_\tj}\;
\dxila p_\tj+\{q^\ti,\dxila \cn^\tsi \gam_\tsi\}.}{7.19}
Thus the transformations $\delta q,\;\delta p$ and $\delta\cn$ should
satisfy the equation \ref{7.19} on the hypersurface $\cm$. Only in this
case can the transformations in phase space be symmetry transformations
of the corresponding Lagrangean system. Substituting \ref{7.14b}
into \ref{7.19} this condition simplifies to
\begin{equation}
{\delta^2(H+\cn^\tsi \gam_\tsi)\over \delta p_i^x\delta p_j^y}EM(p_j^y)\xi^y=
{\delta^2 \gam_\tsi\over \delta p_i^x\delta p_j^y}EM(p_j^y)\lam^\tsi
\label{7.20}
\end{equation}
and imposes a certain functional
dependence between $\xi$ and $\lam^a$. If this condition is
fulfilled the phase space transformations \ref{7.14b} can be interpreted as
Lagrangean symmetries. At the same time the number of free functions
becomes equal to the number of constraints as it should be.\par
\paragraph{Gauge Invariance.} If the constraints are linear
and $H$ at least quadratic in the momenta then only for $\xi^z=0$
can equation \ref{7.20} be satisfied
\footnote{If $H$ and all constraints are at most linear in momenta,
as it is the case for the Chern-Simons theories,
then the Hamiltonian system is strongly degenerate.}.
So, in this case the transformations which are generated by
the constraints alone will also be symmetry transformations for the
corresponding Lagrangean system. We shall call them {\it gauge
transformations}.
For example, in Yang-Mills theories all constraints are linear
in the momenta and as we have seen the
finite gauge transformations can be recovered as transformations
generated only by the constraints ($\xi\es 0$) in the Hamiltonian formalism
\footnote{Another interesting class of theories where all constraints
are linear in momenta are the constrained Wess-Zumino-Novikov-Witten
theories \cite{f92}.}.
\paragraph{Reparametrization invariance.} Usually the reparametrization
invariance of a Lagrangean system, if it exists, is identified with the
gauge invariance in the Hamiltonian formalism.
As we shall see they are actually very different and this identification
can only be made on mass shell.\par
If some of the constraints in are nonlinear in momenta then
it is obvious that the transformations generated by the constraints
only ($\xi=0$) do not satisfy the condition \ref{7.20} and hence can not
be symmetry transformations of the corresponding Lagrangean system.
However, in all known theories with nonlinear constraints
there are canonical coordinates such that $H=0$ and the
condition \ref{7.20} can be satisfied if we impose some functional
dependence between $\lam$ and $\xi$ in \ref{7.14b} so that $\xi\neq 0$
for such theories. Thus the nonlinear constraints generate the
symmetry transformations of the corresponding Lagrangean system only
in very special combination with 'trivial' $\xi$-transformations.
The reason for that is the following: a transformation generated
by a nonlinear constraints takes off mass-shell trajectories away from the
subspace $\cm$ and the extra
compensating transformation returns the trajectories back to $\cm$.
More explicitly taking $\lam^\tsi$ to be $\lam^{e z}=
\cn^{e z}\xi^z$ in \ref{7.20} we reduce this equation to
\begin{equation}
\cn^{e z}(\xi^y-\xi^z)
{\delta^2 \gam_{e z}\over \delta p^{ix}\delta p^{jy}}
EM(p^{jy})=0.\label{7.21}
\end{equation}
One sees at once that if
\begin{equation}
{\delta^2 \gam_{e z}\over \delta p^{ix}\delta p^{jy}}\sim \delta(z-y)
\end{equation}
then even for constraints nonlinear in the momenta the equation
\ref{7.21} is satisfied off mass shell ($EM(p)\neq 0$). From that
it follows immediately that the transformations \ref{7.14b}
with $\lam^{e z}\es\cn^{e z}\xi^z$ are symmetry transformations
for the corresponding Lagrangean system if $H\es 0$.
We shall call the corresponding invariance reparametrization invariance:
$\hr_\xi = \hi_{\xi,\lam^{e z}=\cn^{e z}\xi^z}$. The explicit
form of the reparametrization transformations read
\begin{eqnarray}
\delta_\xi
q^{ix}&=&\dot{q}^{ix}\xi^x+(\xi^y-\xi^x)\{q^{ix},\cn^{b y}\gam_{b
y}\}\nonumber\\
\delta_\xi p_i^x&=&\dot{p}_i^x\xi^x+(\xi^y-\xi^x)\{p_i^x,\cn^{b y}\gam_{b y}\}
\label{7.22}\\
\delta_\xi \cn^{a x}&=&(\cn^{a x}\xi^x)^\cdot-\xi^y\cn^{b y}
\cn^{c z}t^{a x}_{c z,b y}.\nonumber
\end{eqnarray}
These transformations are the correct ones for theories
with non-linear constraints. For example, for
the relativistic particle the transformations \ref{7.22} (and not
the gauge transformations \ref{7.8} generated by the first class
constraints alone) coincide with \ref{7.4} on $\cm$.\par
For the relativistic particle, string and gravity the transformations
\ref{7.22} form a {\it closed algebra off mass shell}. This closure is not ment
to be obvious since it depends on the concrete structure of the constraints.
We verified it for each concrete system separately \cite{mw92}.\par
For the supersymmetric particle there are two constraints
which are quadratic in the momenta and \ref{7.22} is
replaced by a two-parametric family of symmetry transformations.
They are the supersymmetric extensions of the
reparametrization transformations and do not coincide (off
mass shell) with the gauge transformations generated by the constraints.
\par
2An interesting question to which we have no general answer is the following:
what are the conditions to exponentiate the infinitesimal
transformations to finite ones. For the relativistic particle and string
and for gravity the finite transformations for the corresponding
Lagrangean systems are just the familiar symmetries. These
finite symmetries can then be formulated in the Hamiltonian formulation
and this way one can find the finite transformation in the first
order formalism. But in general it is not clear whether the
closing of the algebra of infinitesimal transformations is sufficient
to make them finite. We suppose that this
c2annot be the case since
for a free {\it nonrelativistic particle}, which very probably
does not admit any known finite local
symmetry the transformations \ref{7.14b} with $\lam\es 0$ form a closed
algebra.
This difficult and very important question (i.e. for the functional
integral) what are the conditions such that the transformations
can be made finite needs further investigation.
\paragraph{Constraints and the equations of motion.} There is a
very interesting and non-trivial connection between the equations
of motion $EM(q)\es EM(p)\es 0$ and the constraints $\gam_\tal\es 0$.
Clearly, since $\dot\gam_\tal\es 0$ the classical trajectories
will stay on $\Gamma_c$.
Inversely, in some theories (e.g. gravity) we can get the equations
of motions (or some of them as in string theory) if we only demand that the
constraints are fulfilled for all $t$ (i.e. everywhere) and that the symmetry
transformations do not destroy this property. For example, in gravity
and string theory this means that we demand that the constraints are
valid everywhere and for any choice of spacelike hypersurfaces, because
the symmetry transformations (diffeomorphism transformations) can
be intepreted as a change of foliation of space-time. It is very easy
to arrive at this conclusion using the developed formalism. Let us
consider how the constraints change under the symmetry transformations
\ref{7.14b}:
\eqngrl{
 \dxila \gam_\tal&=&{\delta \gam_\tal\over \delta q^{ix}}\dxila q^{ix}
+{\delta \gam_\tal\over \delta p_i^x}\dxila p_i^x\nonumber}{
&=&{\delta \gam_\tal\over \delta q^{ix}}EM(q^{ix})\xi^x+
  {\delta \gam_\tal\over \delta p_i^x}EM(p_i^x)\xi^x
  +\lambda^\tga t^\tbe_{\tal\tga}\gam_\tbe.}{7.23}
For the known theories the constraints are local in
$q$ and $p$ and involve only space derivatives of $q$ up to second and $p$
up to first order. It follows then that the
structure of the functional derivative of the constraints
have the form
\eqngrl{
{\delta \gam_{a y}\over \delta q^{ix}}&=&A_{ia}\delta(x,y)+B^j_{ia}
{\pa\over \pa y^j}\delta(x,y)+D^{jk}_{ia} {\pa ^2\over \pa y^j \pa y^k}
\delta (x,y)}{
{\delta \gam_{a y}\over \delta p_i^x}&=&E_a^i\delta(x,y)+F^{ij}_a
{\pa\over \pa y^j}\delta(x,y),}{7.24}
where $A,B,\dots$ are functions of $q^y$ and $p^y$. Substituting
\ref{7.24} into \ref{7.23} a straightforward calculation yields
\begin{eqnarray}
\dxila \gam_{a y} &=&\big(\dot{\gam}_{a y}+\cn^\tbe t^\tga_{\tbe ,a y}
\gam_\tga +t^\tga_{a y} \gam_\tga\big)\xi_y\nonumber\\
&+&\lam^\tga t^\tbe_{a y,\tga} \gam_\tbe+\Big(B^j_{ia} EM(q^{iy})+
F^{ij}_a EM(p_i^y)\Big){\pa\xi^y\over \pa y^j}\\
&+&D^{jk}_{ia}\Big(2{\pa EM(q^{iy})\over \pa y^j}{\pa\xi^y\over \pa y^k}
+EM(q^{iy}){\pa^2\xi^y\over \pa y^j\pa y^k}\Big).\nonumber
\end{eqnarray}
It follows that for systems with a {\it finite number of
degrees of freedom} $\delta \gam\sim \gam$, so that if we impose
the constraints for a phase-space path then the transformed
path satisfies also the constraints. For that none of the
equations of motion must be satisfied. Thus for a finite number
of degrees of freedom the constraints $\gam_a$ do not tell us anything
about the dynamics of the system.\par
Also, the gauge transformations ($\xi=0$) do respect the constraints
since $\delta_{\lam ,\xi=0}\gam\sim \gam$ as it is clear from \ref{7.23}.
Again we see that the constraints together with the corresponding
symmetry transformations tell us nothing about the
dynamics.\par
However, for field theories with non-linear constraints
we must take $\xi\neq 0$ in \ref{7.14b}. In this case if not all
of the coefficients $B,D$ and $F$ vanish we can get (some of) the
equations of motion by demanding that the symmetry transformations
respect the constraints, i.e. that $\delta \gam\es 0$ if $\gam\es 0$
everywhere. Whether some of the coefficients $B,D,F$ are non-zero depends
on the concrete structure of the constraints. Only if the constraints
contain some space derivatives of $q$ and/or $p$ are some of the
coefficients non-zero. For example, in {\it string theory} only $B\neq 0$
and correspondingly only some equations of motion, namely
two of the $EM(q)\es 0$ equations can be gotten by demanding that all
constraints should
be satisfied everywhere for any foliation. In {\it gravity} the situation
is even more interesting, since all of the coefficients are non-zero
and thus {\it all} equations of motion will be
automatically satisfied if we demand just that $\gam_\tal\es 0$ for any
foliation. So the whole dynamics in gravity can be reduced
to the requirement
that the constraints are invariant under diffeomorphism transformations.

\end{document}